\documentclass[5p,twocolumn]{elsarticle}
\usepackage{graphics,graphicx}
\usepackage{mathrsfs}
\usepackage{amsmath,amsfonts,amssymb,wasysym}
\usepackage{color}
\usepackage{natbib}
\usepackage{epstopdf}
\usepackage{epsfig}
\usepackage{bm}
\usepackage{listings}
\usepackage{graphicx}
\usepackage{newtxtext}
\usepackage{newtxmath}
\usepackage{hyperref}
\usepackage{subcaption}
\usepackage{comment}
\usepackage{cuted}

\biboptions{longnamesfirst,sort&compress}

\hypersetup{
    colorlinks = true,
    urlcolor   = blue,
    citecolor  = black,
}

\newcommand{\RomanNumeralCaps}[1]
\linenumbers

\def\half{\mbox{$1\over2$}}

\newcommand{\bs}{\boldsymbol}
\newcommand{\ucd}{\stackrel{\kern0.1em\triangledown}}
\newcommand{\lcd}{\stackrel{\kern0.1em\triangle}}
\newcommand{\coro}{\stackrel{\kern0.1em\circ}}

\def\tc{{\check\tau}}

\def\cD{{\cal D}}

\def\cL{{\cal L}}

\def\cG{{\cal G}}
\def\cF{{\cal F}}
\def\cH{{\cal H}}
\def\cU{{\cal U}}
\def\di{{\rm d}}
\def\dgam{{\dot{\gamma}}}

\def\th{{\hat{t}}}

\def\ph{{\hat{p}}}

\def\xh{{\hat{x}}}
\def\yh{{\hat{y}}}
\def\uh{{\hat{u}}}

\def\Ub{{U}}

\def\vh{{\hat{v}}}

\def\cT{{\cal T}}
\newcommand{\be}{\begin{equation}}
\newcommand{\ee}{\end{equation}}
\newcommand{\pd}[2]{\frac{\partial #1}{\partial #2}}
\newcommand{\pdd}[2]{\frac{\partial^2 #1}{\partial #2^2}}

\def\half{{\mbox{$1\over2$}}}

\def\quar{{\mbox{$1\over4$}}}
\def\Bi{{\rm Bi}}

\def\ctxx{{\check\tau_{\rm xx}}}
\def\ctxy{{\check\tau_{\rm xy}}}
\def\ctyy{{\check\tau_{\rm yy}}}
\def\ttxx{{\tilde\tau_{\rm xx}}}
\def\ttxy{{\tilde\tau_{\rm xy}}}

\def\txx{{\tau_{\rm xx}}}
\def\txy{{\tau_{\rm xy}}}
\def\sxx{{\sigma_{\rm xx}}}
\def\sxy{{\sigma_{\rm xy}}}
\def\syy{{\sigma_{\rm yy}}}
\def\Txx{{T_{\rm xx}}}
\def\Txy{{T_{\rm xy}}}
\def\tyy{{\tau_{\rm yy}}}
\def\txxh{{\hat\tau_{\rm xx}}}
\def\txyh{{\hat\tau_{\rm xy}}}
\def\tyyh{{\hat\tau_{\rm yy}}}

\def\Yb{{\bar{Y}}}
\def\Yc{{\check{Y}}}

\def\ttxx{{\tilde\tau_{\rm xx}}}
\def\ttxy{{\tilde\tau_{\rm xy}}}
\def\tpsi{{\tilde\psi}}

\def\vxx{{\varsigma_{\rm xx}}}
\def\vxy{{\varsigma_{\rm xy}}}
\def\hpsi{{\hat\psi}}
\def\ve{{\varepsilon}}

\journal{Journal of Non-Newtonian Fluid Mechanics}

\begin{document}

\begin{frontmatter}

\title{Start-up and inertialess instability of elasto-viscoplastic channel flow}

\author[ad1]{J. D. Shemilt\corref{cor1}} 
\author[ad1]{N. J. Balmforth} 
\author[ad2]{D. R. Hewitt}

\address[ad1]{ Department of Mathematics, University of British Columbia,
  Vancouver, BC, V6T 1Z2, Canada}
    \address[ad2]{Department of Applied Mathematics and Theoretical Physics,
University of Cambridge, Wilberforce Road, Cambridge CB3 0WA, UK}
\cortext[cor1]{Corresponding author: {\it E-mail:} shemilt{@}math.ubc.ca}

\date{\today}

\begin{abstract}
  An exploration is presented of
the start-up and linear stability of
pressure-driven channel flow
of an elasto-viscoplastic fluid described by Saramito's
constitutive law.
Streamwise uniform base states are non-unique, depending
on the initial stress configuration, and
   develop discontinuities in the normal stresses
   and shear rate at the yield surfaces over infinite times.
   Such stress discontinuities can be eliminated by introducing a
   sufficient extensional pre-stress; true plugs bordered
   by stress jumps then become replaced by marginally yielded,
   plug-like flow, or pseudo-plugs. To examine the 
   stability of all of these states, the linear initial-value
   problem is solved along with the evolving base states. Because
   this analysis is performed for finite times, the base
   states remain continuous and there is no
   need to perturb any stress discontinuities.
   Armed with the insights provided, stability is then
   analyzed as a normal-mode problem for the final
   states, building in perturbations
   to the stress discontinuities {\it via} certain jump conditions
   across any yield surfaces. Regardless of whether the
   base flows contain true plugs or pseudo-plugs, the base states
   are found to be linearly unstable at zero Reynolds number.
   The most unstable perturbations possess the highest streamwise
   wavenumbers and become spatially localized to the regions
   where stresses lie close to the yield stress. 
\end{abstract}

\end{frontmatter}

\numberwithin{equation}{section}

\section{Introduction}

In an effort to understand the impact of richer rheology on
the dynamics of yield-stress fluids, a number of recent studies 
have revisited classical problems using elasto-viscoplastic constitutive
models. In these efforts, Saramito's model \cite{saramito07}
has proved popular. The key feature of this constitutive law is the
introduction of a switch that allows the Oldroyd-B model
of viscoelasticity to be combined with a Bingham-type yield stress
fluid model (or, after further elaboration, the Herschel-Bulkley
law \cite{saramito09}). Recent efforts have thereby quantified
the effect of elasticity on viscoplastic flow around
spheres and rising or oscillating bubbles, the thinning of viscoplastic
filaments and spreading drops
\cite{frag,mosch,franca2024,zakeri25,moschopoulos23,chaparian19,chaparian20}.

The goal of the present paper is to reconsider the pressure-driven flow
of an elasto-viscoplastic fluid through a channel. This
non-Newtonian version of the
Poiseuille flow problem has received limited attention,
despite its classical nature, both as a
start-up and a linear stability problem.
Here, we reconsider both facets, examining the consequences of
employing Saramito's model to describe the constitutive behaviour.

In the start-up problem, and in line with previous discussion of
non-uniqueness of other simple flows \cite{cheddadi12},
we first demonstrate that the steady states reached 
depend on the initial stress conditions.
We next establish that the constitutive law
generically leads to the formation of discontinuities in the normal
stresses and shear rate at the yield surfaces
in the final steady states.
The emergence of stress discontinuities in circular Couette
flow has previously been discussed by Cheddadi {\it et al.}
\cite{cheddadi08,cheddadi12}.
Here, we also demonstrate that the
stress jumps can be eliminated if flow
is initiated with a sufficient pre-stress, albeit at
the expense of creating extensive, weakly yielded
plug-like regions, or ``pseudo-plugs'' following the
terminology of Walton \& Bittleston \cite{walton_axial_1991}.
Armed with insight into the range of possible
final states that can be reached in the start-up problem,
we then advance to an investigation of linear stability,
explicitly considering the inertialess limit.

In other non-Newtonian contexts, stress discontinuities in a steady base
state can act as the seed for linear instability,
even at zero Reynolds number
(\textit{e.g.}, \cite{chen91,hinch92,castillo22,Fielding05}).
Inertialess viscoelastic shear flows are also the setting for a
number of other kinds of instabilities 
(see the reviews \cite{castillo22,datta22}). Notably, Wilson and co-workers
have explored an instability that occurs for 
power-law polymer viscosities that are strongly shear-thinning
\cite{wilson99,wilson15,castillo18,bodiguel,poole16}.
Indeed, Patne \cite{patne}
recently exploited the analogy between a yield stress and such power-law
viscosities to suggest that fully-yielded Couette flow of an
elasto-viscoplastic fluid can be unstable at zero Reynolds number.
Here, however, we identify a different instability in inertialess channel flow 
that is intrinsically linked to the unyielded plug or pseudo-plug of
the base state for a fluid described by Saramito's model.

When the base state features stress discontinuities,
significant complications arise in dissecting the character of this linear
instability. In particular,
it is not immediately clear how to introduce perturbations
to the stress jumps at the yield surfaces.
We take two approaches to resolve this situation.
First, we solve the linear initial-value problem while concurrently
evolving the start-up base state. Because the base-state stress solution
is then continuous at any finite time, we avoid the need to
consider yield surfaces with stress jumps.
The results of the linear initial-value problem then guide us
in reformulating a normal-mode analysis of the final
discontinuous base state, applying certain jump conditions
across the yield surfaces. The normal-mode analysis is simpler when the
base state instead contains a pseudo-plug and is continuous.
We find that the structure of the steady base state
critically impacts the linear stability properties.



The remainder of the paper is organised as follows. After presenting the model
equations in \S\ref{sec:formu}, we explore the base states in \S\ref{sec:base},
and interrogate the impact of varying the initial pre-stress.
In \S\ref{sec:stab}, we then explore
the linear stability of a number of possible base states.
We conclude with a discussion of our results in \S\ref{sec:discu}.  

\section{Mathematical formulation\label{sec:formu}}

\subsection{Governing equations}

We consider an elasto-viscoplastic fluid flowing down a two-dimensional
channel of width $\cH$
under the action of a mean pressure gradient $-\Gamma$.
We use Cartesian coordinates $(\xh,\yh)$ to describe the
geometry; the flow field is $(\uh,\vh)$.
Discarding inertia, the incompressibility and force balance
conditions demand
\begin{align}
  \pd{\uh}{\xh} &+ \pd{\vh}{\yh}=0, 
  \\
0 &= \pd{\sxx}{\xh} + \pd{\sxy}{\yh} ,\\
0 &= \pd{\sxy}{\xh} + \pd{\syy}{\yh} , 
\end{align}
where $\bs\sigma$ is the total stress tensor.
At the walls of the channel, we impose the no-slip conditions,
\be
\uh=\vh = 0 \qquad {\rm at} \ \yh=\pm\half\cH.
\ee

Saramito's model \cite{saramito07}
splits up the stress into polymer and solvent
components, so that
\be
\bs{\sigma} = -\ph \bs{I} + \mu_s \hat{\dot{\bs{\gamma}}} + \bs{\hat{\tau}},
\label{estress}
\ee
where $\ph$ is a pressure, $\mu_s$ is the solvent viscosity
and the strain rates are given by
\be
\hat{\dot{\bs{\gamma}}} = \left(\begin{matrix}
  \mbox{\scriptsize 2} \pd{\uh}{\xh} & \pd{\uh}{\yh}+\pd{\vh}{\xh} \\
  \pd{\uh}{\yh}+\pd{\vh}{\xh} & \mbox{\scriptsize 2} \pd{\vh}{\xh}
\end{matrix}\right)
.
\ee
The constitutive law is
\be
\lambda \ucd{\bs{\hat\tau}} + Y \bs{\hat\tau} = \mu_p \hat{\dot{\bs{\gamma}}},
\ee
where $\lambda = \mu_p/G$ is the relaxation time,
the triangular diacritic denotes the upper convected derivative,
$G$ is the elastic modulus, $\mu_p$ is the polymer viscosity, and
\be
Y = \text{max} \left\{ 0, 1 - \frac{\tau_Y}{\hat\tau_d } \right\},
\ee
is Saramito's switch, with $\hat\tau_d$ representing
the second invariant of the deviatoric part of $\bs{\hat\tau}$.



\subsection{Dimensionless model equations}

To non-dimensionalise, we introduce scalings
based on the channel width $\cH$,
the dimensional pressure gradient driving flow $\Gamma$,
and the sum of the polymer and solvent viscosities, $\mu_p+\mu_s$.
We further use a different length-scale $\cL$ to characterize flow
perturbations in the streamwise direction, and adopt
different scales for the components of the velocity and stress.
These choices are
guided by our interest in the high Weissenberg limit
of the constitutive model, where elastic effects are particularly pronounced. A similar high Weissenberg limit has been exploited to fabricate lubrication theory for Oldroyd-B-type
fluids without a yield stress \cite{zhang02,ahmed21,hinch24,boyko24}.
Specifically, we take
\be
\cL = \lambda \cU , \qquad
\delta = \frac{\cH}{\lambda\cU} = \frac{\cH}{\cL},
\qquad
\cU = \frac{\Gamma \cH^2 }{ 2(\mu_p+\mu_s)}.
\label{scal1}
\ee
The velocity scale $\cU$ corresponds to that 
generated for a viscous or Oldroyd-B fluid
with total viscosity $\mu_p+\mu_s$, given the 
imposed pressure gradient. Therefore, the choice of length scale $\cL$
implies that the aspect ratio, $\delta$, is
equivalent to an inverse Weissenberg number. Connecting the length scale $\cL$
and elastic properties in this way allows for a convenient simplification
of the model equations in the long-wave, high-Weissenberg-number limit
$\delta\ll1$. However, we state the equations below for general $\delta\geq0$,
and also investigate the model for $\delta>0$ as well as the simpler case
with $\delta=0$. 

We now set
\be
\th = \frac{\cL}{\cU} t,\quad
\xh = \cL x, \quad \yh = \cH y,
\ee
\be
(\uh,\vh) = \cU (u,\delta v), \quad 
(\hat\dgam,\hat\dgam_{ij}) = \frac{\cU}{\cH} (\dgam,\dgam_{ij}),
\ee
\be
(\ph,\txxh,\txyh.\tyyh) = \half\Gamma\cL
\left(p - 2x,\txx,\delta\txy,\delta^2\tyy  \right)
.
\label{scal2}
\ee
The dimensionless model equations 
then become
\begin{align}
\pd{u}{x} &+ \pd{v}{y}=0, \label{conteq}\\
\pd{p}{x} &=
\beta\left(\pdd{u}{y} + \delta^2 \pdd{u}{x}\right)
+ \pd{\txx}{x} + \pd{\txy}{y} + 2,\\
\pd{p}{y} &=
\beta \delta^2\left(\pdd{v}{y}+ \delta^2\pdd{v}{x}\right)
+ \delta^2 \left(\pd{\txy}{x}+\pd{\tyy}{y}\right), \label{1.9}\\
  \frac{D\txx}{Dt} &+ Y \txx
  - 2\txx\pd{u}{x} - 2\txy\pd{u}{y}
  = 2\delta^2(1-\beta)\pd{u}{x},   \label{1.11}
  \\
  \frac{D\txy}{Dt} &+ Y \txy
- \txx\pd{v}{x} - \tyy\pd{u}{y}
= (1-\beta)\left(\pd{u}{y}+ \delta^2 \pd{v}{x}\right),
\\
  \frac{D\tyy}{Dt} &+ Y \tyy
- 2\txy\pd{v}{x} - 2\tyy\pd{v}{y}
= 2(1-\beta)\pd{v}{y}.
  \label{1.13}
\end{align}
where
\be
\frac{D}{Dt} = \pd{}{t} + u \pd{}{x} + v \pd{}{y} 
,
\qquad
Y = \max\left(1-\frac{\Bi}{\cT},0\right)
\ee
and
\be
\cT = \sqrt{(\txx-\delta^2\tyy)^2+4\delta^2\txy^2}.
\ee
Here,
\be 
\beta = \frac{\mu_s}{\mu_s+\mu_p}  \quad {\rm and} \quad
\Bi = \frac{2\tau_Y\cH^2}{(\mu_s+\mu_p) \cU \cL}
\ee
denote a viscosity ratio and Bingham number, respectively. 
The other main parameters in the model are $\delta$, which is the inverse Weissenberg number as defined in \eqref{scal2}, and a dimensionless perturbation wavenumber that we impose when considering the linear stability of the base states. The initial stress conditions must also be specified for the base state; we introduce a fifth parameter controlling the initial extensional pre-stress in the following section.

\section{Base states\label{sec:base}}


The model equations admit a solution describing
uniform flow down the channel in which
\be
u = \Ub(y,t), \quad
\txx = \Txx(y,t), \quad
\txy = \Txy(y,t), 
\ee
with $\tyy=v=0$. This base-state solution satisfies
\begin{align}
\beta \pdd{\Ub}{y} + \pd{\Txy}{y} &= -2, \label{315} \\
  \pd{\Txx}{t} + Y \Txx &= 
  2 \Txy\pd{\Ub}{y}  ,   \label{1.17n}
\\
\pd{\Txy}{t} + Y \Txy &= (1-\beta)\pd{\Ub}{y}.
\label{1.18n}
\end{align}
Assuming that the shear rate $\partial U/\partial y$ and stress $\Txy$ are
anti-symmetric about the channel's centerline,
\eqref{315} implies that
\be
\pd{\Ub}{y} = - \frac{\Txy+2y}{\beta}.
\label{2.7}
\ee
In combination with \eqref{1.17n}-\eqref{1.18n}, we now arrive at
\begin{align}
  \pd{\Txx}{t} + \Yb \Txx &= -\frac{2 }{\beta} \Txy (\Txy+2y)
  \label{baseq1}
\\
 \pd{\Txy}{t} + \Yb \Txy &=
- \frac{(1-\beta)}{\beta} (\Txy+2y)
,
\label{baseq2}
\end{align}
where
\be
\Yb = \left(
1 - \frac{\Bi}{\cT}  \right)  \Theta(\cT-\Bi),
\qquad
\cT = \sqrt{\Txx^2+4\delta^2\Txy^2},
\label{321}
\ee
and $\Theta(X)$
is Heaviside's step function.

If the fluid begins unyielded and remains so, then $Y=0$
and we find the solution,
\begin{align}
  \Txy &=
  \left[\Txy(y,0)+2y\right]\exp \left[-\frac{(1-\beta)t}{\beta}\right]
  - 2y
  \label{3.23}
, \\
\Txx &= \frac{[\Txy(y,t)]^2 - [\Txy(y,0)]^2}{1-\beta} 
+ \Txx(y,0)
,
  \label{3.24}
\end{align}
with $\Ub(y,t)$ decreasing exponentially with time.
Hence, the stresses evolve towards the motionless steady state,
\be
\begin{aligned}
  \Txx &= (1-\beta)^{-1}\left\{4y^2-\left[\Txy(y,0)\right]^2\right\}
  + \Txx(y,0), \\
\Txy &= -2y,\qquad
  \Ub = 0.
\end{aligned}
\label{2.14}
\ee

If, however, the yield stress becomes breached at some stage
due to the growth of the stresses, the evolution
is more complicated with the base state evolving towards
a final profile given algebraically by
\be
\begin{aligned}
  1 - \frac{\Bi}{\sqrt{\Txx^2+4\delta^2\Txy^2}}
  &= -\frac{(1-\beta)(\Txy+2y)}{\beta \Txy} ,\\
  \Txx &= \frac{2 \Txy^2}{(1-\beta)} .
\end{aligned}
\label{2.17}
\ee
This final state
is independent of the initial stress configuration
aside from the requirement that the fluid becomes yielded
at an earlier time,
which is partly dictated by $\Txx(y,0)$ and $\Txy(y,0)$ through
\eqref{3.23} and \eqref{3.24}. Note that $\delta$ appears in these base-state solutions only via $\cT$; the only impact of increasing $\delta$ on the base state is, thus, modifying the yield criterion, and it has little qualitative impact on the structure of the base flow. 

Depending on initial conditions, other evolutionary pathways are possible
in which fluid begins yielded, then becomes unyielded, or
yields at finite time before becoming unyielded at late times.
Below, we present specific examples illustrating a range
of potential base states.

\subsection{No initial polymer stress}\label{sec:sampsolsSARA}

\begin{figure*}[ht!]
  \centering
	\includegraphics[width=.95\linewidth]{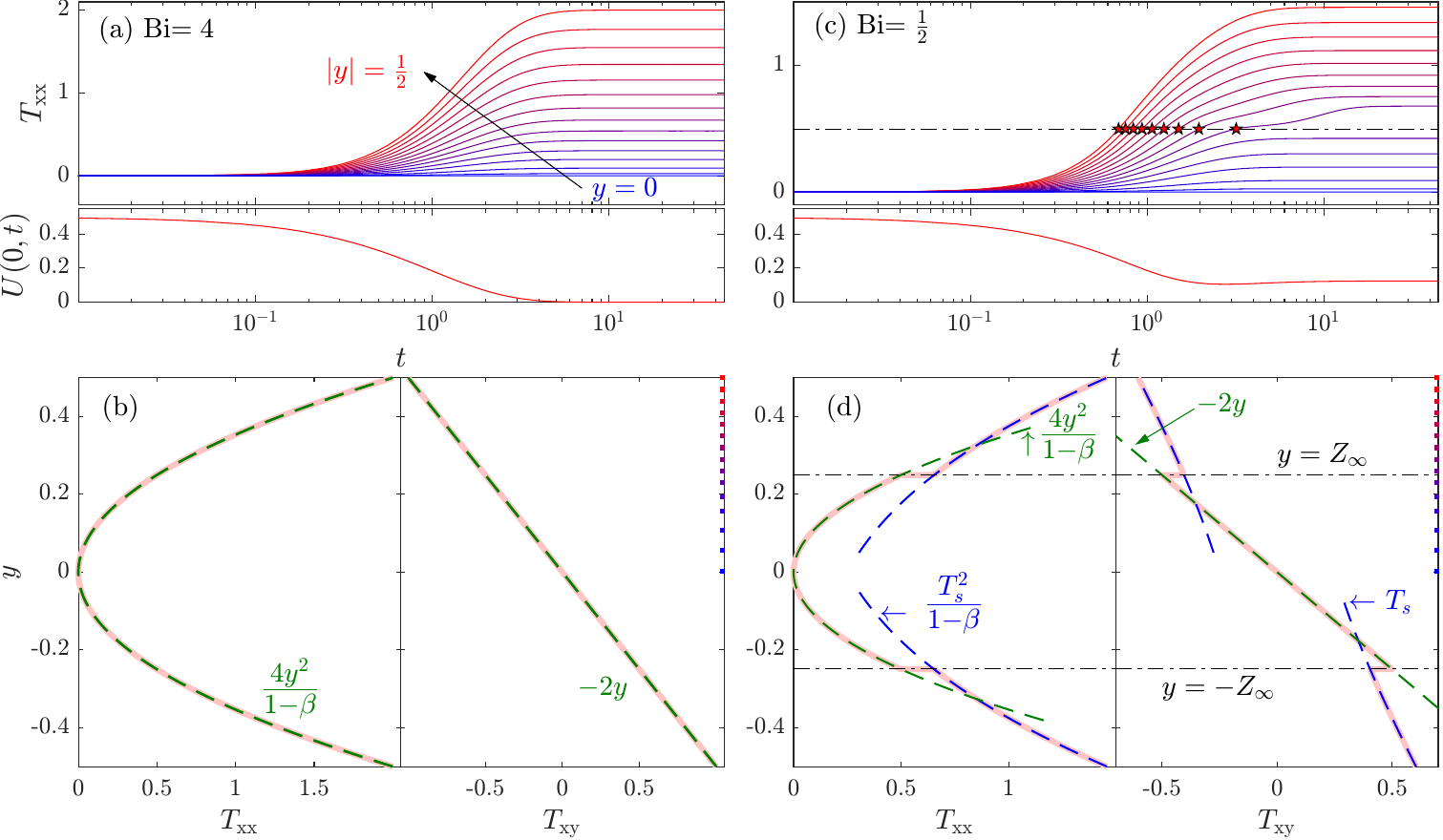}
	\caption{Base state solutions with no initial polymer stress, 
          for (a,b) $\Bi=4$, which evolves to a motionless final state, and (c,d) $\Bi=\frac12$, which evolves to a flowing final state with a central plug. In both cases, 
          $\{\delta,\beta\}=\{\frac{1}{20},\frac12\}$
          and $\Txx(y,0)=\Txy(y,0)=0$.
          Plotted in (a,c) are times series of $\Txx(y,t)$ 
          at fifteen stations in $y$ across the upper
          half of the channel (with the solutions
          colour-coded, from blue to red and as indicated
          by the arrow, on progressing
          away from the centre of the channel),
          and the central speed $U(0,t)$.
          The stars in (c) indicate
          the moments at which the yield stress becomes breached
          at that particular spatial position.
          The dot-dashed line in (c) is the approximate yield threshold, $\Txx=\Bi=\tfrac{1}{2}$. 
          In (b,d), we plot the final profiles of
          $\Txx$ and $\Txy$ (thicker, solid pink lines).          
          The dashed lines indicate
          the steady state profiles from \eqref{2.14} and \eqref{2.17a}
          (green and blue, respectively), and the dot-dashed
          lines show the levels where the yield condition
          is met for the final state ($y=\pm Z_\infty$).
          The squares on the right of (b,d) indicate the positions at which
          $\Txx$ is plotted in (a,c).
}
	\label{fig:baspoiplot}
\end{figure*}

For the special case $\Txx(y,0)=\Txy(y,0)=0$, the fluid is initially
unyielded and
\begin{align}
  \Txy &= -2y \left[ 1 -  e^{-(1-\beta)t/\beta}\right]
, \\
  \Txx &= \frac{4 y^2}{1-\beta}
\left[ 1 -  e^{-(1-\beta)t/\beta}\right]^2
.
\end{align}
These stress components reach their largest values for $y=\pm\frac12$.
Hence, if
\be
\Bi > \Bi_* = \frac{\sqrt{1+4\delta^2(1-\beta)^2}}{1-\beta},
\label{Bc}
\ee
the yield stress can never be breached, and the solution
converges to a motionless elastic final state in which
\begin{align}
  \Txy = -2y \qquad {\rm and} \qquad
  \Txx &= \frac{4 y^2}{1-\beta} .
  \label{2.14a}
\end{align}
Figure \ref{fig:baspoiplot}(a,b) presents a corresponding numerical solution
of the initial-value problem.
Note that $\Bi_*\approx(1-\beta)^{-1}$, if $\delta\ll1$.

By contrast,
if $\Bi<\Bi_*$ in \eqref{Bc},
the yield condition is met at some time along the walls of the channel.
Yield surfaces then migrate into the fluid from the walls.
Nevertheless, because $\Txx=\Txy=0$ at $y=0$ for
any time with this
initial condition, the yield criterion is never met along the
channel's centerline,
and an unyielded core always remains. The core becomes bounded
by yield surfaces that eventually approach the levels,
$y=\pm Z_\infty$, where
\be
\Bi = 4 Z_\infty \sqrt{\delta^2+\frac{Z_\infty^2}{(1-\beta)^2}}
  .
  \label{2.19}
  \ee

Figure \ref{fig:baspoiplot}(c,d) presents a second numerical solution
in which $\Bi=\frac12< \Bi_*$. In this example,
the yield stress becomes breached near the walls shortly
after $t=1$. Now, over the yielded regions that develop
against the walls, the final state is given instead by
\eqref{2.17}.
Note that the solutions shown in figure \ref{fig:baspoiplot}
are computed with the relatively small value $\delta=\frac{1}{20}$,
and for $\delta\ll1$, the final yielded profile
in \eqref{2.17} simplifies to
\be
\begin{aligned}
  \Txy &\approx T_s \equiv -(1-\beta)\left[{\rm sgn}(y)\sqrt{y^2
      + \frac{\beta\Bi}{2(1-\beta)}} + y\right] , \\
  \Txx &= \frac{2\Txy^2}{(1-\beta)} .
\end{aligned}
\label{2.17a}
\ee
Crucially, 
it is clear from \eqref{2.14a} and \eqref{2.17a}
that the stress profiles of the two final states 
cannot connect continuously across the final yield surfaces $y=\pm Z_\infty$
in \eqref{2.19}.
In particular, the stress components take the values
\be
\Txx
\approx \Bi , \qquad 
\Txy
\approx \pm2Z_\infty \approx \pm\sqrt{(1-\beta)\Bi} ,
\ee
on the unyielded side
of the yield surfaces, but
\be
\begin{aligned}
\Txx&\approx
\half\Bi\left(1-\beta+\sqrt{1+\beta^2}\right)^2,  \\
\Txx&\approx
\pm Z_\infty\left(1-\beta+\sqrt{1+\beta^2}\right),
\end{aligned}
\ee
on the yielded side. The 
stress components must therefore develop jump discontinuities across the
yield surfaces. This development can be observed in
figure \ref{fig:baspoiplot}(c) by the divergence of the trajectories
of $\Txx$ to either side of the yield condition $\Txx\approx\Bi$.
The stress jumps are also evident in
figure \ref{fig:baspoiplot}(d), which compares the numerically computed
final stress profiles with the predictions
in \eqref{2.14} and \eqref{2.17}. 

Although $\Txx$ and $\Txy$ become discontinuous for $t\to\infty$,
the stress profiles remain continuous for any finite time.
The manner in which $\Txx$ and $\Txy$ converge to their final
discontinuous profiles
is examined more thoroughly in \ref{sec:steps}. There, we show that
the profiles steepen exponentially quickly, and determine a local
self-similar coordinate change that maps out the
structure of the steepening steps in $\Txx$ and $\Txy$. 
Note also that,
although the polymeric shear stress, $\Txy$, develops a discontinuity,
the total shear stress, $\Txy + \beta U_y=-2y$,
is always continuous, whereas the shear rate $U_y$ develops a similar
jump to $\Txy$. (Final profiles of $U$ for each of the sample solutions presented in this section are illustrated in figure \ref{fig:baspoiplot4} below.)


  \begin{figure*}[t!]
  \centering
	\includegraphics[width=\linewidth]{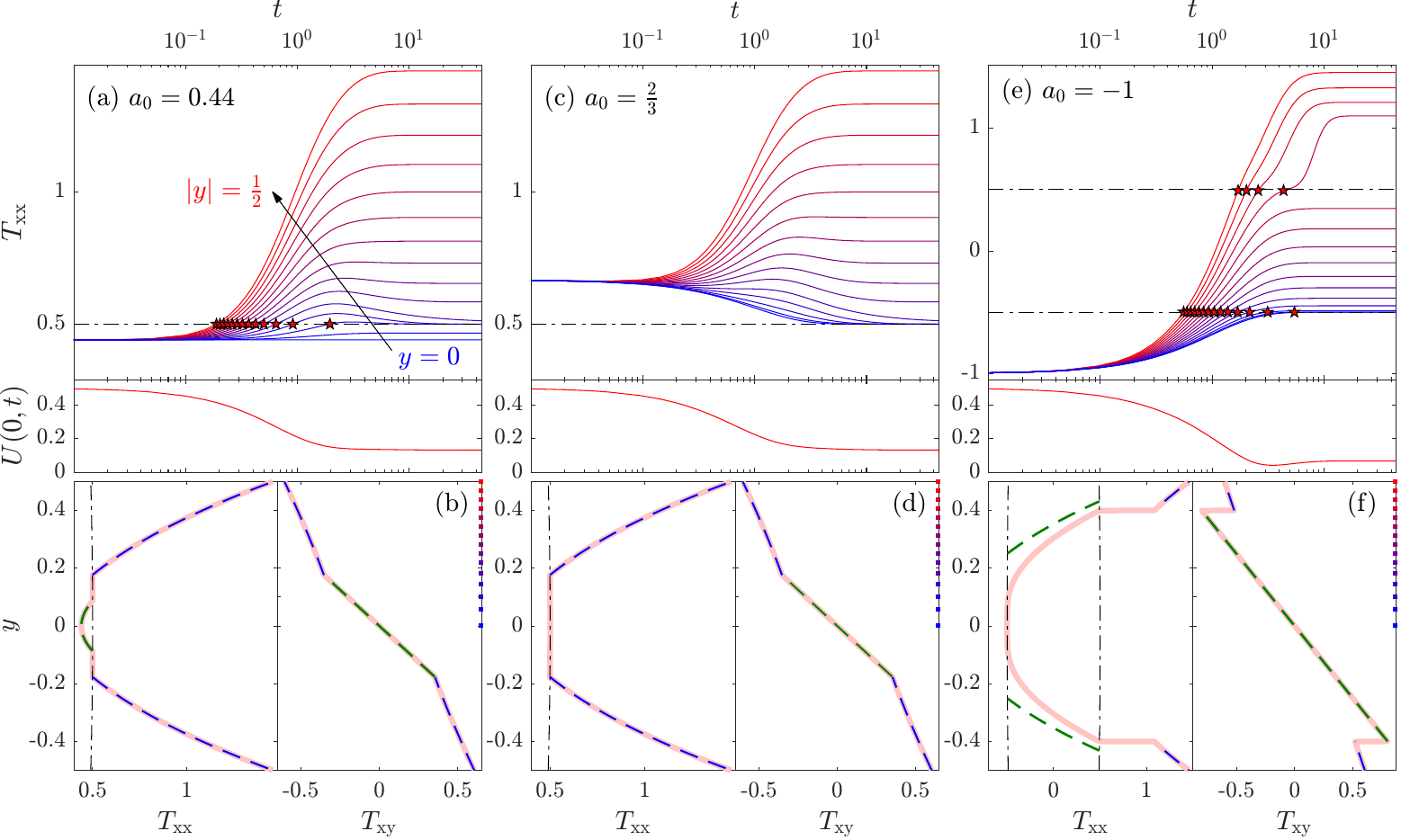}
	\caption{Base state solutions for $\Bi = \frac12$ with initial extensional pre-stress
          (a,b) $a_0=0.44$, (c,d) $a_0=\frac23$ and (e,f) $a_0=-1$, and with
          $\{\delta,\beta\}=\{\frac{1}{20},\frac12\}$,
          $\Txx(y,0)=a_0$ and $\Txy(y,0)=0$.
          Plotted are (a,c,e) times series of $\Txx(y,t)$ for
          fifteen stations in $y$ across the upper
          half of the channel (colour-coded by position $y$) and $U(0,t)$, 
          and (b,d,f) the final profiles of
          $\Txx$ and $\Txy$ (thick pink lines). In (a,c,e), stars indicate
          the moments at which the yield stress becomes breached
          at each value of $y$, and dot-dashed lines indicate the approximate yield thresholds $\Txx=\pm\Bi$.
          The dashed lines in (b,d,f) indicate
          the steady state profiles from \eqref{2.14} and \eqref{2.17a}
          (green and blue, respectively), and the dot-dashed
          lines are the final-state yield thresholds for $\Txx$. 
          The squares indicate the positions at which
          $\Txx$ is plotted in (a,c,e).
}
	\label{fig:baspoiplot2a}
  \end{figure*}

  \subsection{Evolution with extensional pre-stress}
  \label{sec:sampsolsSARAb}

When different initial conditions are adopted, the dynamics
described above becomes more varied. Solutions taking
$\Txy(y,0)=0$ and $\Txx(y,0)=a_0$, for constant $a_0$, are illustrated in
figure \ref{fig:baspoiplot2a}. 
This initial condition, with a uniform extensional pre-stress,
is not selected with any specific method of preparation in mind
(and may be difficult to set in any practical setting), but
provides a relatively simple choice that generates a range of start-up flows. 

When $a_0$ is relatively small and positive, the main effect of the
extensional pre-stress $a_0$
is to shift the final profile of $\Txx$ to the right,
as indicated by \eqref{2.14}. The yield surfaces and their
stress jumps become displaced accordingly, but the dynamics largely
follows the same pattern as that seen in figure \ref{fig:baspoiplot}(c,d).
When $a_0$ is larger, however,
the stress relaxes viscoelastically 
over a fraction of the layers that yield at intermediate times;
see figure \ref{fig:baspoiplot2a}(a,b). This relaxation leads to
the stress declining back to the yield stress from
above at later times. The stress does not pass back through
that threshold, however, but converges to it for $t\to\infty$.
This behaviour is permitted by the constitutive equations
in \eqref{baseq1}-\eqref{baseq2}
because the time derivatives in \eqref{baseq1} and \eqref{baseq2}
converge to zero for
\be
\Txx\to \pm\sqrt{\Bi^2 - 4\delta^2y^2}\approx \pm \Bi, \quad
\Txy\to -2y ,
\quad
Y\to0.
\label{2.16}
\ee
The regions described by \eqref{2.16} are yielded for any finite
time, but have exponentially small shear rates with
a stress invariant $\cT\approx\Bi$. This feature leads us to refer
to these regions as
``pseudo-plugs'', following Walton \& Bittleston's
terminology for Bingham fluids \cite{walton_axial_1991}. 
\begin{figure*}[h!]
  \centering
	\includegraphics[width=0.9\linewidth]{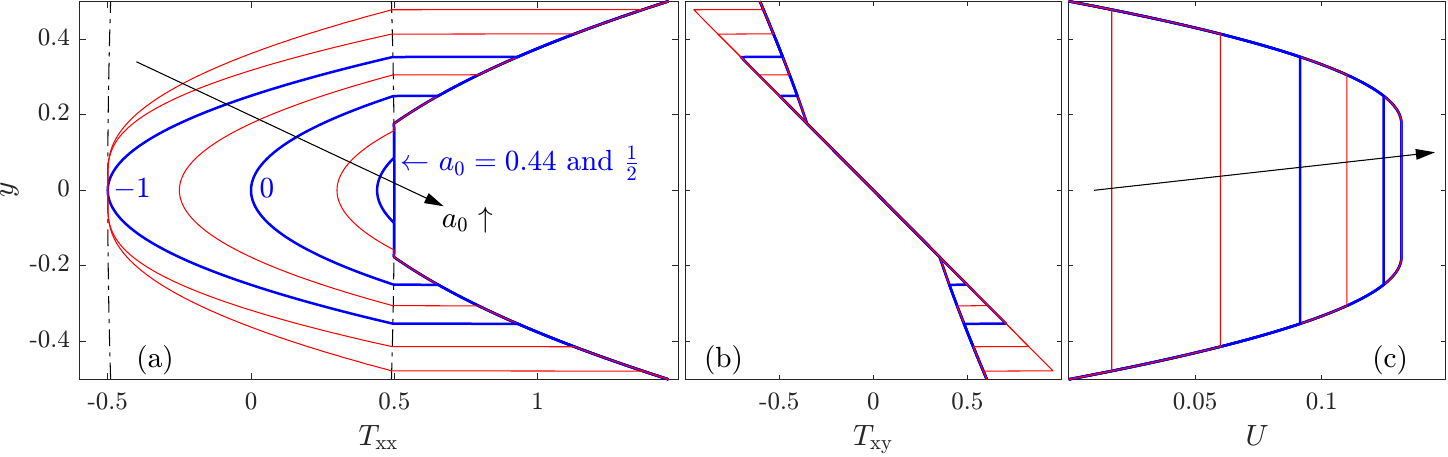}
	\caption{Final profiles of (a) $\Txx$, (b) $\Txy$
          and (c) $U$, with
          $\{\Bi,\delta,\beta\}=\{\frac12,\frac{1}{20},\frac12\}$,
          $\Txx(y,0)=a_0$ and $\Txy(y,0)=0$, 
          for $a_0=\{-20,-2,-0.8,-0.3,0,0.25,0.5,0.7,1 \}$.
          The cases shown by thicker (blue)
          lines indicate the examples shown 
          in figures \ref{fig:baspoiplot} and \ref{fig:baspoiplot2a}.
}
	\label{fig:baspoiplot4}
\end{figure*}

For the example shown in figure \ref{fig:baspoiplot2a}(a,b),
the extensional pre-stress is not sufficient to
breach the yield condition over the centre of the channel.
The final state then contains a central pseudo-plug
with a true plug embedded within it.
For even higher $a_0\geq\Bi$,
as in the example shown in figure \ref{fig:baspoiplot2a}(c,d),
the initial pre-stress is sufficient to yield
the entire layer at the outset. The central core then remains
yielded for all time, with the 
fluid there converging to the pseudo-plug flow in \eqref{2.16}.
The final state then pieces together yielded buffers
at the walls with the central pseudo-plug, at least provided
$(1-\beta)\Bi\lesssim 1$ (as in figure \ref{fig:baspoiplot2a}(c,d)).
If that yield criterion is violated,
however, but $a_0 \geq \Bi \gtrsim (1-\beta)^{-1}$, the initial state is
yielded and relaxes to a motionless state in which
$(\Txx,\Txy)\to(\Bi,-2y)$ and the pseudo-plug
fills the entire channel.

For compressive pre-stresses, $a_0<0$, the final profile of
$\Txx$ shifts to the left, but as long as $a_0>-\Bi$ the dynamics
remains similar to that in figure \ref{fig:baspoiplot}(b).
Taking $a_0<-\Bi$, however, again leads to a fully yielded initial
state, as illustrated by the example shown in
figure \ref{fig:baspoiplot2a}(e,f).
The extensional stress now increases with $t$
throughout the entire fluid layer, with the stress
approaching the threshold
$\Txx = -\sqrt{\Bi^2 - 4\delta^2y^2}\approx -\Bi$ for times
of order unity. The fluid then becomes unyielded
and begins to deform elastically, first near the walls, but
only over long times near the centre of the channel.
For the solution in figure \ref{fig:baspoiplot2a}(e,f),
the stress near the walls eventually increases
sufficiently to breach the other yield threshold
$\Txx = \sqrt{\Bi^2 - 4\delta^2y^2}\approx \Bi$ at even later times.
As shown by the final stress profiles in figure \ref{fig:baspoiplot2a}(f),
the central plug then becomes
buffered from the walls by yielded regions.
Note that, in this example,
the final stress profile of $\Txx$ is no longer given
by \eqref{2.14} over the unyielded regions owing to the unrecoverable
plastic deformations taking place during the initial period when
the fluid there was yielded. 

For other choices of the parameters ($a_0$,$\Bi$,$\delta$,$\beta$), yet more
qualitatively different final states can be reached.
A selection of final profiles for a suite of solutions
with varying $a_0$, holding the other parameters
fixed, is shown in figure \ref{fig:baspoiplot4}. The final panel
of this figure displays the associated velocity profiles.
The cases shown by thicker blue lines indicate the examples shown earlier
in figures \ref{fig:baspoiplot} and \ref{fig:baspoiplot2a}. 

Two further noteworthy cases occur when
fluid is initially yielded by taking
either sufficiently large extensional or compressive pre-stress,
but the final state is rendered motionless
by a suitably chosen yield stress 
(neither of which are illustrated in
figures \ref{fig:baspoiplot}-\ref{fig:baspoiplot4}).
In the former case, the extensional pre-stress yields the fluid
initially if $a_0>\Bi$, but the stresses then relax towards
the state $(\Txx,\Txy)\approx(\Bi,-2y)$ everywhere
provided the yield stress exceeds the threshold $\Bi>\Bi_*$;
{\it i.e.} the fluid evolves to a motionless steady state
with a pseudo-plug filling the entire channel.
In the compressional case, situations can arise in which
taking $a_0<-\Bi$, with both $|a_0|$ and $\Bi$ sufficiently large,
leads to the entire fluid layer reaching the yield threshold
$\Txx\approx-\Bi$ at intermediate times, but then failing
to breach the threshold $\Txx\approx\Bi$ at the wall
at late times (unlike in figure \ref{fig:baspoiplot2a}(e,f)).
In other words, fluid motion can become suppressed
altogether in the steady state by a compressional pre-stress
even when the yield stress is below the threshold ($\Bi < \Bi_*$).

Allowing for different initial, spatially varying pre-stresses may
lead to a yet richer variety of final stress states.
Below, in considering linear stability,
we focus on base states generated via start-up from a stress-free initial state, such as that shown in figure \ref{fig:baspoiplot}(c,d), and base states with pseudo-plugs like those in figure \ref{fig:baspoiplot2a}(a-d).
The latter are potentially relevant when fluid is prepared by imposing a much stronger transient
pressure gradient that briefly yields most of the fluid. After
that ``pre-pressuring'', if the
pressure gradient is reduced back to the level $\Gamma$
to allow fluid stresses to relax back towards the yield stress, pseudo-plugs
are likely to appear in the centre of the channel.


\section{Linear stability\label{sec:stab}}

We test the linear stability of the base states 
by considering infinitesimal wave-like perturbations
described by a streamwise wavenumber $k$ and streamfunction:
\be
\begin{aligned}
  p&=\check{p}(y,t)e^{ikx},\\
  [\txx,\txy] &= [\Txx(y,t),\Txy(y,t)] +
   [\ctxx(y,t),\ctxy(y,t)]e^{ikx},\\
  \tyy &= \ctyy(y,t)e^{ikx},\\
[u,v] &= [\Ub,0] + 
   \left[-\psi'(y,t),ik\psi(y,t)\right]e^{ikx} ,\qquad\qquad
   \label{2.19n}
\end{aligned}
\ee
where the prime signifies differentiation
with respect to $y$.
Then, after a linearization of \eqref{conteq}-\eqref{1.13}, we find
\begin{align}
  ik \check{p} = ik\ctxx + \ctxy' - &\beta (\psi''' - \delta^2 k^2 \psi') ,
  \label{1.16}
\\
\check{p}' = \delta^2 [\ctyy' + ik \ctxy &+ ik\beta (\psi'' - \delta^2 k^2\psi)] ,\\
  \cD \ctxx  + \Txx \Yc -
  2 \Ub' \ctxy
  &=  
  - ik \Txx' \psi 
  - 2ik \Txx \psi' 
  - 2  \Txy \psi'' \nonumber \\
  \ & \qquad - 2ik\delta^2 (1-\beta) \psi'
  \label{1.17}
    , \\
\cD \ctxy + \Txy \Yc
  - \Ub' \ctyy   
  &= - (1-\beta) (\psi'' + \delta^2k^2\psi)
- ik \Txy' \psi \nonumber \\
  \ & \qquad - k^2\Txx \psi
  \label{1.18}
  , \\
 \cD \ctyy 
  = 2ik(1-\beta) \psi' 
  &- 2k^2\Txy \psi,
  \label{1.19}
\end{align}
where
\be
\cD = \pd{}{t} + ik\Ub + \Yb ,
\ee
the base-state switch in \eqref{321} is again written as $\Yb$,
and
\be
\Yc = \frac{\Bi}{\cT^3} \left[\Txx (\ctxx-\delta^2\ctyy)
+ 4 \delta^2 \Txy \ctxy\right] \Theta(\cT-\Bi)
.
\ee
The boundary conditions on the walls become
\be
\psi(\pm\half)=\psi'(\pm\half)=0 ,
\ee


Equations \eqref{1.16}-\eqref{1.19} can be solved as an initial-value
problem given an evolving base state like those explored in
\S\ref{sec:sampsolsSARA} and \$\ref{sec:sampsolsSARAb}.
Alternatively, we can consider
the final steady base states and perform a standard normal-mode-type
analysis, adopting the time dependence $e^{\sigma t}$, where
$\sigma=\sigma_r+i\sigma_i$ is a (complex) growth rate.
A key issue with the latter approach, however, are the
stress discontinuities at the yield surfaces, which must be perturbed
in some way. Equivalently, in \eqref{1.17}-\eqref{1.18}, 
we must interpret the coefficients involving $\Txx'$ and $\Txy'$
at $y=\pm Z_\infty$.

To begin, we therefore focus on the initial-value problem,
beginning from initial conditions with
\begin{eqnarray}
	\ctxx(y,0) = 1,\quad \ctxy(y,0)=\ctyy(y,0)=0,
\end{eqnarray}
when computing solutions in which $\psi$ is an even function
(indicating that $\ctxy$ is odd), or
\begin{eqnarray}
	\ctxy(y,0) = 1,\quad \ctxx(y,0)=\ctyy(y,0)=0,
\end{eqnarray}
when computing solutions for which $\psi$ is odd. In practice, we solve
the initial value problem in the half-domain, $-\tfrac{1}{2}\leq y\leq0$,
imposing symmetry or antisymmetry conditions, as appropriate, at $y=0$. 
In this setting, the base-state stresses remain
continuous but sharpen with time in the vicinity of the yield surfaces
({\it cf.} \ref{sec:steps}).
To further simplify matters, we base our exploration
on the long-wave, very visco-elastic limit of the model equations,
$\delta\ll1$. \ref{appA} outlines the simplifications
offered by taking this limit, and how these may be incorporated
into a scheme for numerically solving the model equations.

After exploring the initial-value problem, we then consider
base states without stress discontinuities for which
a traditional normal-mode analysis is possible. In this setting, we
follow a numerical strategy in which we
turn the derivatives in $y$ in \eqref{1.16}-\eqref{1.19} into centred
differences on a finite grid or use Chebyshev differentiation
matrices (we implemented both options). Either way, this device
recasts the linear stability equations as a matrix eigenvalue
problem for the stress components after introducing
the dependence $e^{\sigma t}$. Given an eigenvalue/eigenvector
pair computed in this manner,
the problem can then be re-solved using a boundary-value solver
with an adaptive grid to refine the solution (we use Matlab's
BVP4C). As we outline presently
and discuss more thoroughly in \ref{sec:jumpy},
the normal-mode problem can also
be generalized for discontinuous base states. However,
the numerical strategy needed in such cases is more involved,
leading us to provide a briefer study of the
normal modes of discontinuous base states. 

\subsection{Linear initial-value calculations}\label{sec:ivp}

\begin{figure}[t!]
  \centering
	\includegraphics[width=\linewidth]{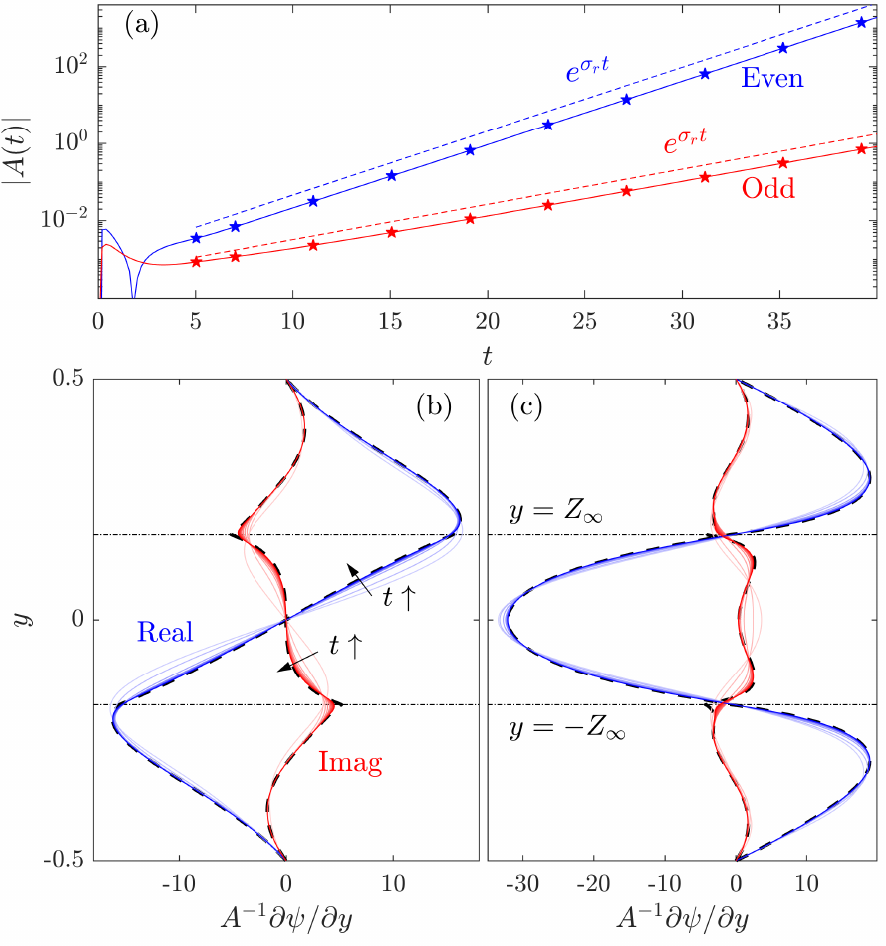}
	\caption{
          Solutions of the
          linear initial-value computation for
          even and odd modes, with
          $k=10$ and $\Bi=\beta=a_0=\frac12$.
          Plotted are 
          (a) time series of $|A(t)|$ and
          (b,c) snapshots of the real and imaginary parts
          of $A^{-1} \psi_y$ at the times indicated by stars in (a).
          The dashed lines show
          the exponential growth $e^{\sigma_r t}$
          and spatial profiles of
          the most unstable normal modes.
          The dot-dashed
          lines indicate the yield positions $y=\pm Z_\infty$.
        }
	\label{fig:ivpit}
\end{figure}

\begin{figure}[t!]
  \centering
	\includegraphics[width=\linewidth]{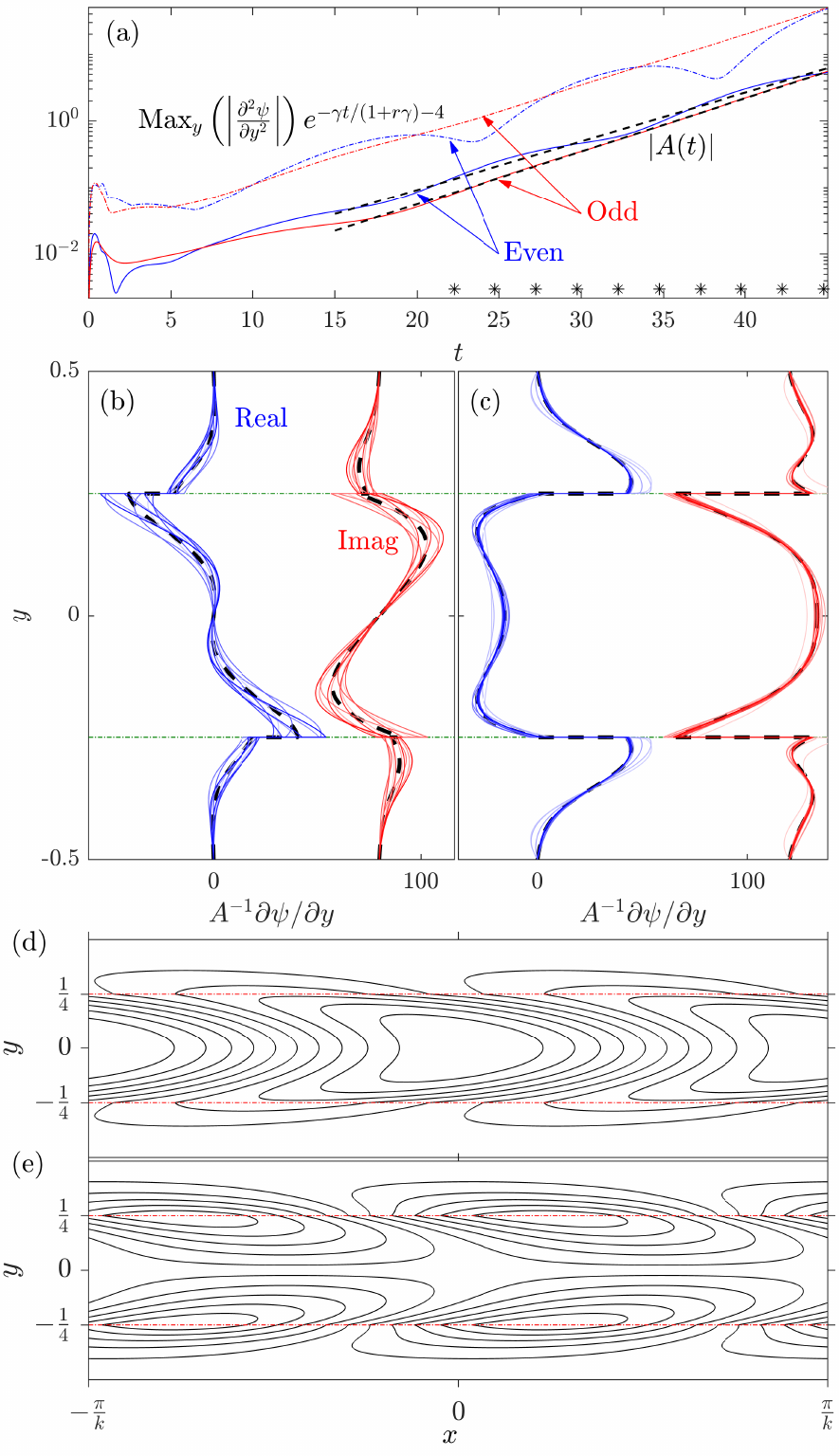}
	\caption{
          Solutions of the
          linear initial-value computation for
          even and odd modes, with
          $(k,\Bi,\beta,a_0)=(20,\frac12,\frac12,0)$.
          In (a) we show time series of $A(t)$ in \eqref{AA}
          (solid lines)
          and the maximum over $y$ of
          $\left|\partial^2\psi/\partial y^2\right|$, scaled by
          a factor of $\ve(t)=e^{-\gamma t/(1+r\gamma)}$.
          The even mode is shown by darker (blue) lines;
          the odd mode by lighter (red) lines.
          The dashed lines show exponential fits to the
          time series of $A(t)$.
          In (b,c), we plot
          snaphots of 
          $A^{-1}\partial\psi/\partial y$
          at the times indicated by the stars in (a),
          for the even and odd modes, respectively.
          Snapshots are plotted darker as time increases, with the
          real parts on the left and the imaginary parts offset
          to the right. The dashed lines
          show corresponding
          unstable normal-mode profiles, computed as described
          in \ref{sec:jumpy}. In (d,e), we show contours of constant
          Re$[e^{ikx}\psi(y,t=45)]$ for the two initial-value solutions.
          The dot-dashed
          lines indicate the yield surfaces $y=\pm Z_\infty$.
        }
	\label{fig:plo2}
\end{figure}

\begin{figure}[t!]
  \centering
	\includegraphics[width=\linewidth]{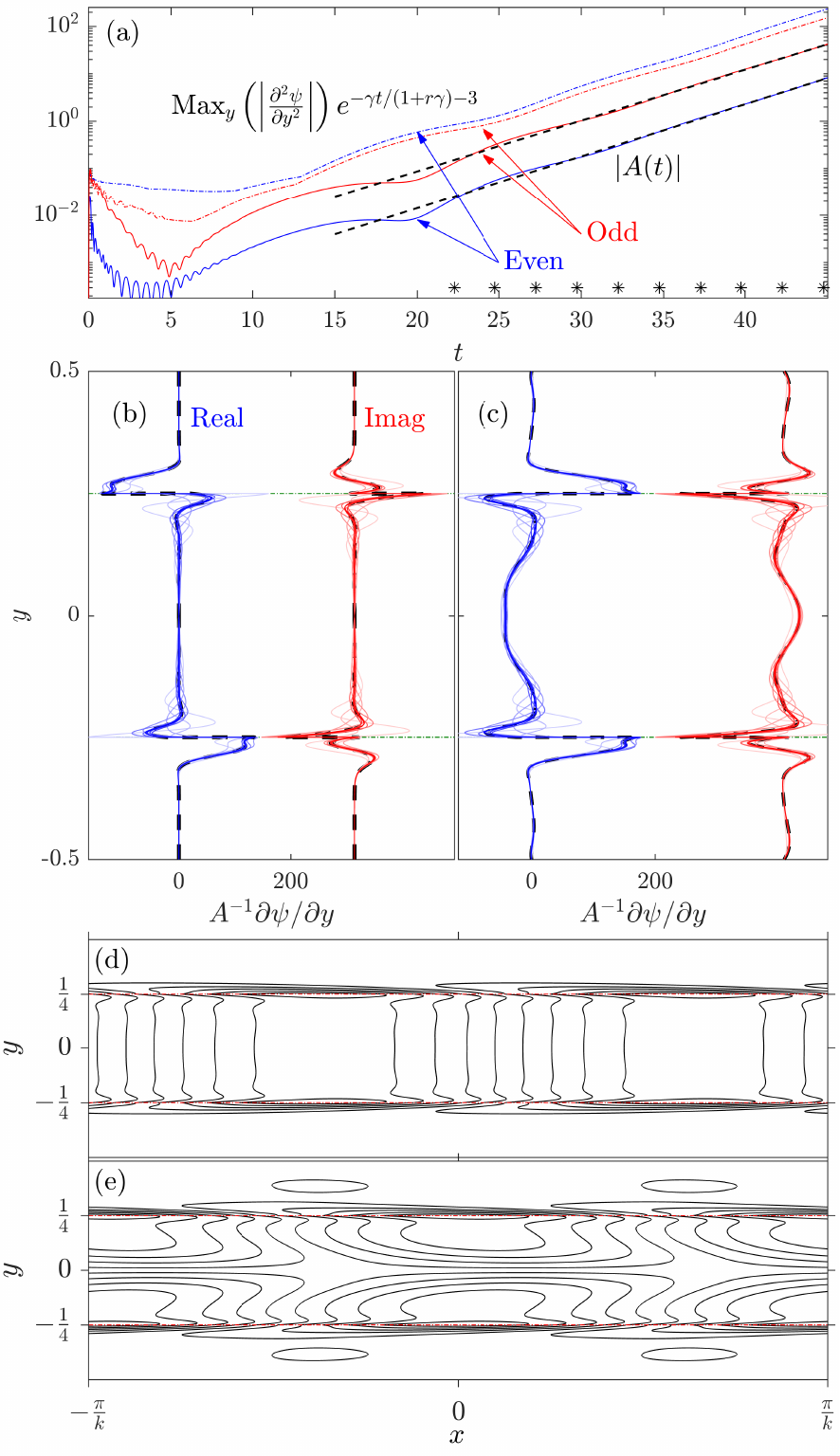}
	\caption{
          A similar set of plots to those in figure \ref{fig:plo2},
          but for $k=100$.
        }
	\label{fig:plo1}
\end{figure}

Solutions to the initial-value problem for two different base states
are shown in figures \ref{fig:ivpit}, \ref{fig:plo2} and \ref{fig:plo1}.
In the first figure, the base state corresponds to that
shown in figure \ref{fig:baspoiplot2a}(c,d) (except that $\delta=0$);
fluid is fully yielded by an extensional
pre-stress at $t=0$, and develops a central pseudo-plug
at later times. Presented is a
pair of even or odd solutions in $\psi(y,t)$,
showing time series of the perturbation amplitude
$|A(t)|$, defined using
\be
A(t) = \int_{0}^{1/2} \psi\;\di y,
\label{AA}
\ee
and snapshots of the scaled streamwise velocity perturbation,
$A^{-1}\partial\psi/\partial y$.
For both the even and odd solutions, the mode amplitude $|A(t)|$
follows an initial transient for times $t<10$, which corresponds
to the interval over which the base state progresses
close to its final steady profile (see figure \ref{fig:baspoiplot2a}(c)).
Thereafter, $|A(t)|$ converges to exponential
growth, whilst $A^{-1}\partial\psi/\partial y$ approaches a fixed
spatial structure, implying a convergence to 
a normal-mode form. Indeed, solving the normal-mode
problem directly for the most unstable modes
furnishes the exponential growth and eigenfunctions shown by
dashed lines in the plots of figure \ref{fig:ivpit}.
We consider the normal-mode problem
in more detail in \S\ref{sec:nm} below, in order to more completely survey the
character of the linear stability.
For the moment, however, we return to the initial-value
problem and consider a base state in which the stress profiles
sharpen towards discontinuities at $y=\pm Z_\infty$.


The initial-value solutions
shown in figures \ref{fig:plo2} and \ref{fig:plo1}
are computed
using the base state of figure \ref{fig:baspoiplot}(c,d)
(but with $\delta=0$).
The two figures again present pairs of even or odd solutions,
this time for two different wavenumbers.
Once more, the time series of the perturbation amplitude $|A(t)|$
shown in the top panels
expose what appear to be the emergence of exponentially
growing modes. However, the situation
is now more obscure: also plotted is the
maximum over $y$ of $\left|\partial^2\psi/\partial y^2\right|$, scaled by
$\ve(t)=e^{-\gamma t/(1+r\gamma)}$, where
\be
\gamma = \frac{\sqrt{1+4\beta-4\beta^2}-1}{2\beta}
  \qquad {\rm and} \qquad
  r = \frac{1-\beta}{\beta}
  .
\ee
The scaled shear-rate perturbation grows exponentially
at a similar rate to $|A(t)|$. By contrast
true normal modes possess the pure
exponential time-dependence $e^{\sigma t}$ and should
require no further temporal scaling. Thus,
the initial-value solution for $\partial^2 \psi/\partial y^2$
cannot have true normal-mode form. This feature
carries over to the stress perturbations
$\ctxx(y,t)$ and $\ctxy(y,t)$, whose maximum absolute values
require the same temporal scaling.

The origin of this additional scaling lies in
the linear equations \eqref{1.16}-\eqref{1.19}, which
contain coefficients involving the spatial
gradients of $\Txx$ and $\Txy$. As we demonstrate
in \ref{sec:steps} for base states that develop stress discontinuities,
the base-state stress gradients sharpen exponentially quickly with a
factor $\ve^{-1}=e^{\gamma t/(1+r\gamma)}$. Evidently, this
exponential sharpening becomes translated
into the linear perturbations. In fact,
the scaled streamwise velocity perturbation
$A^{-1}\partial\psi/\partial y$ 
develops abrupt jumps like the base-state
stress components (see figures \ref{fig:plo2}(b,c)
and \ref{fig:plo1}(b,c)). Simultaneously, the stress perturbations
and $\partial^2\psi/\partial y^2$
develop sharp, localized peaks that narrow and grow with time
relative to their structure over the rest of the channel.
Corresponding fine structure appears in the contour plots
of Re$(e^{ikx}\psi(y,t))$ presented
in figures \ref{fig:plo2}(d,e) and \ref{fig:plo1}(d,e),
which provide a more global impression of the spatial pattern
associated with the instabilities at $t=45$.

\begin{figure}[t!]
  \centering
	\includegraphics[width=\linewidth]{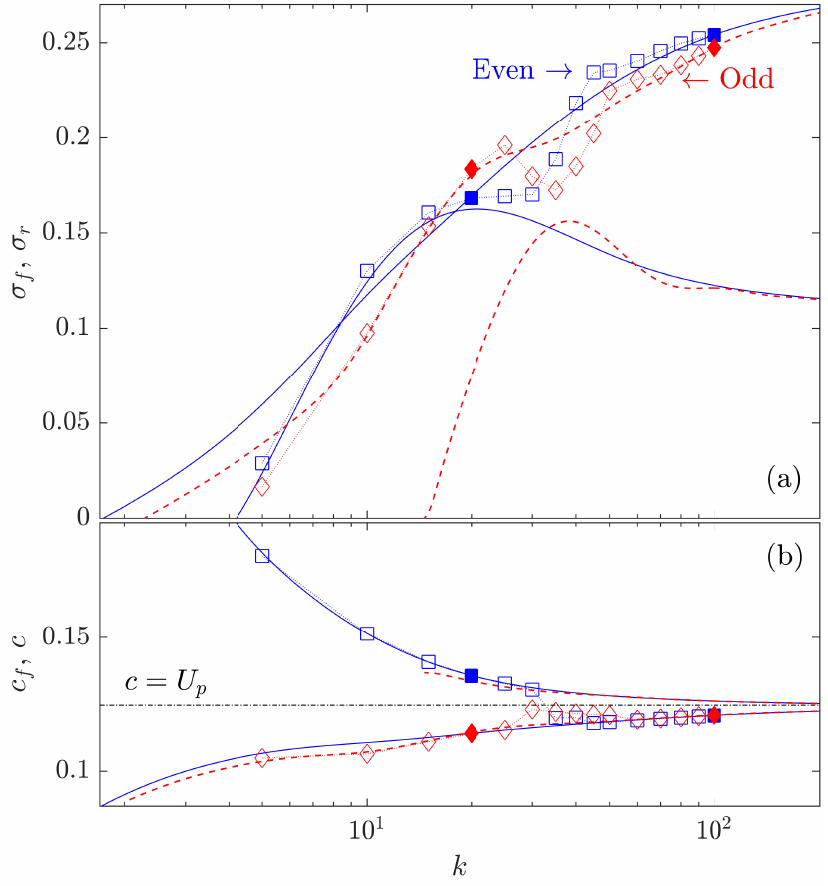}
	\caption{ 
          (a) Growth rates and (b) phase speeds
          over a range of wavenumbers $k$
          for the base state with
          $\Bi=\beta=\frac12$ and $a_0=0$.
          The squares and diamonds present the results,
          $\sigma_f$ and $c_f$, extracted
          from exponential fits
          to the time series of the amplitude and phase of $A(t)$
          from linear initial-value problems          
          ({\it cf.} the dashed lines in figures
          \ref{fig:plo2}(a) and \ref{fig:plo1}(a);
          the filled symbols here indicate these two examples).
          Even solutions are shown by (blue) squares, odd solutions
          by (red) diamonds. The solid blue and dashed red lines
          show the corresponding results, $\sigma_r$ and $c=-\sigma_i/k$,
          for the two pairs of (even and odd)
          normal modes computed using the generalized
          linear stability analysis of \ref{sec:jumpy}.
          The dot-dashed line in (b) indicates the plug speed $U_p$.
        }
	\label{fig:plosum}
\end{figure}

The linear instability exposed by the solution of the
initial-value problem
for the sharpening base state of figure \ref{fig:baspoiplot}
is quantified further in figure \ref{fig:plosum}.
The squares and diamonds shown in this
plot summarize growth rates $\sigma_f$ extracted from
exponential fits to the time series of $|A(t)|$
for a suite of initial-value computations
for even and odd perturbations with varying
wavenumber (the dashed lines
in figures \ref{fig:plo2}(a) and \ref{fig:plo1}(a)
show the fits for $k=20$ and $k=100$).
Figure \ref{fig:plosum}(b) also presents results from fits to the phase
of $A(t)$. That phase decreases linearly with time once the perturbations
begin to grow exponentially, emphasizing how the perturbations
take the form of amplifying travelling waves.
The fits provide a frequency $\omega_f$ that can be converted
to a phase speed $c_f=-\omega_f/k$, which is plotted in panel (b).

The growth rates $\sigma_f$
in figure \ref{fig:plosum} mostly increase with wavenumber,
with sufficiently long waves becoming stable. For $k\gg1$, the phase speed
$c_f$ approaches the speed of the plug $U_p$
({\it i.e.} the maximum of $U(y)$). At these high wavenumbers,
the growth rates and phase speeds
of the even and odd modes also approach one another.
A glance at figure \ref{fig:plo1}(b,c), with $k=100$, exposes the
origin of this convergence: for higher wavenumbers,
the perturbations become increasingly localized to
the yield surfaces. This localization, and the decay of the
perturbations over the intervening plug, 
ensures a relatively weak coupling
between the two sides of the channel.
The spatial structure of the even and odd perturbations then becomes
similar, aside from some more minor differences across the plug
(which are emphasized by
the streamfunction perturbation contours in figure \ref{fig:plo1}(d,e)).
The spatial concentration of the perturbations and the
convergence of the phase speed to $U_p$ (figure \ref{fig:plosum}(b))
highlight how the instability is associated with the regions surrounding the
yield surfaces. 

The impact of the exponential sharpening of the
base-state stress upon the long-time initial-value solutions 
in figures \ref{fig:plo2} and \ref{fig:plo1}
can be understood in more detail from a
generalization to the linear stability problem
of the analysis presented in \ref{sec:steps}.
This generalization is provided in \ref{sec:jumpy}. In brief, the sharpening
steps force linear perturbations to develop a multi-scale
structure: over the bulk of the channel, where the base state
becomes steady, perturbations adopt a global normal-mode-like
form. In the vicinity of the yield surfaces,
however, the continued sharpening of the stress jumps introduces
a time-dependent finer spatial scale with
$y\pm Z_\infty = O(\ve)$.
The stress perturbations and $\partial^2\psi/\partial y^2$
become enhanced over the finely scaled regions
by the factor $\ve^{-1}$, accounting for
the additional scaling required in
figures \ref{fig:plo2}(a) and \ref{fig:plo1}(a)
for Max$_y(\partial^2\psi/\partial y^2)$.
The matching of the local solution near $y=\pm Z_\infty$
to the globel mode over the bulk of the channel
further demands that the latter 
satisfies effective jump conditions
across the yield surfaces. Taking those jumps into consideration
along the lines outlined in \ref{sec:jumpy} leads to
a generalization of the normal-mode analysis. It follows from this
analysis that the global mode adopts
the normal-mode-like dependence $e^{\sigma t}$, even though the
solution does not take a standard normal-form spatial form
over the finely scaled regions.

\begin{figure*}[h!]
  \centering
	\includegraphics[width=.75\linewidth]{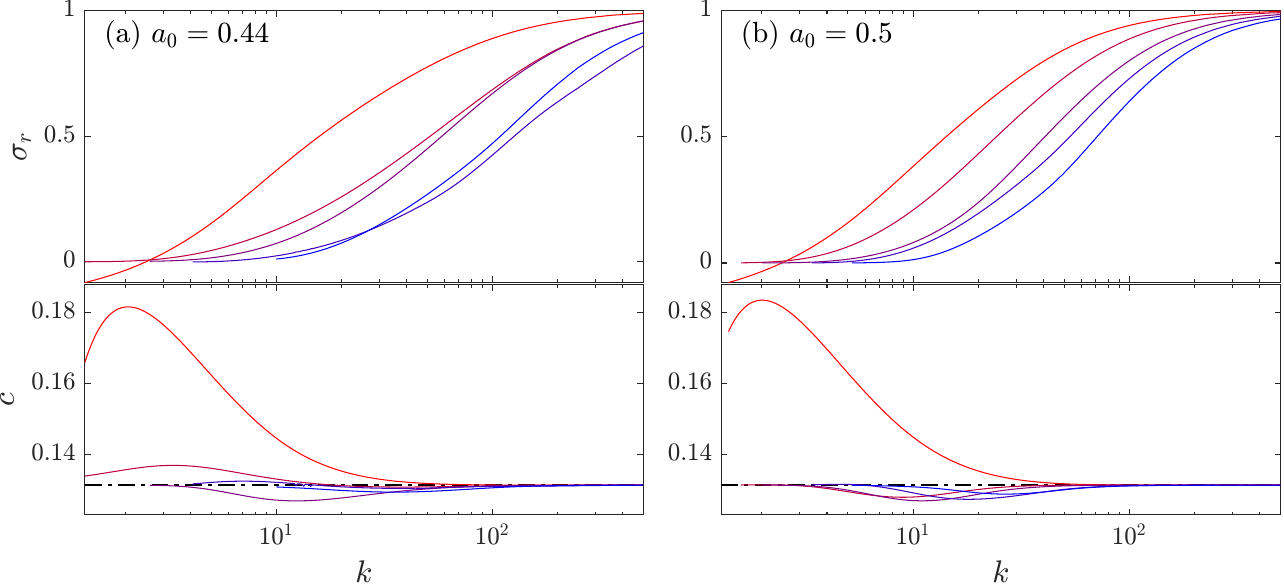}
	\caption{
          The five most unstable normal modes at $k=10$
          for base states with two values of $a_0$
          (as indicated)
          and $(\Bi,\delta,\beta)=(\frac12,0,\frac12)$,
          continued to both higher and lower wavenumbers,
          plotting growth rate $\sigma_r$ and
          phase speed $c=-\sigma_i/k$.
          The dot-dashed lines in the bottom row indicate the
          plug speed $U_p$. 
        }
	\label{fig:sig1plot}
\end{figure*}

\begin{figure*}[t!]
  \centering
	\includegraphics[width=.75\linewidth]{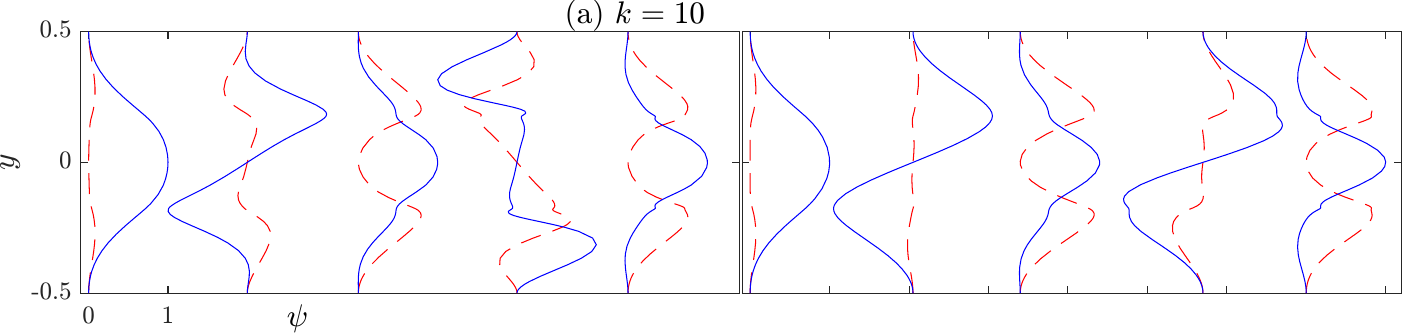}
	\includegraphics[width=.75\linewidth]{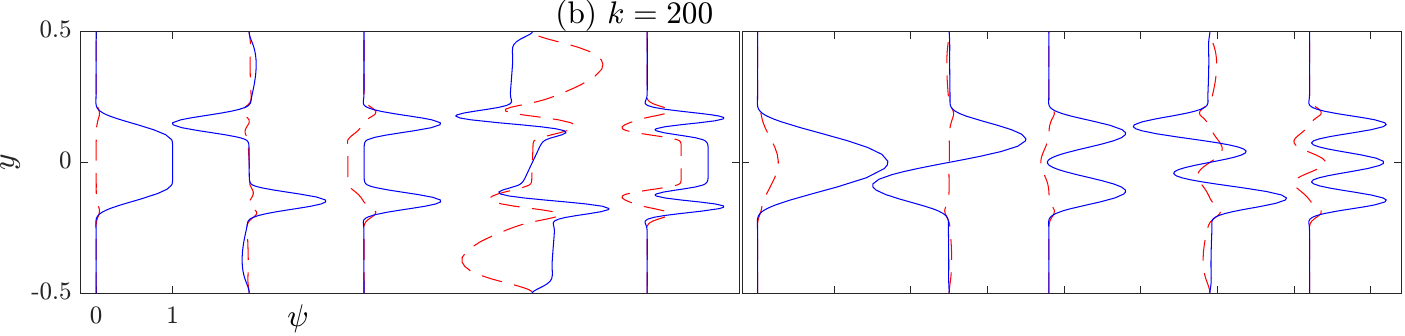}
	\caption{
          Streamfunctions of the normal modes from
          figure \ref{fig:sig1plot} at (a) $k=10$ and (b) $k=200$
          (real part in solid blue,
          imaginary part in dashed red).
          The left and right panels in each row show, respectively, the modes computed with the
          two different base states ($a_0=0.44$ and $a_0=0.5$),
          and the modes are successively offset horizontally
          (starting with the most unstable mode on the left).
        }
	\label{fig:sig3plot}
\end{figure*}

As shown in figures \ref{fig:plo2}(c) and \ref{fig:plo1}(b,c),
the most unstable modes predicted by the generalized normal-mode problem
(included as the dashed lines)
match the spatial structure of the corresponding initial-value solutions.
Both the global-mode structure away from the yield surfaces
and the effective jumps across them are reproduced. Predictions for
the growth rate and phase speed for varying $k$ are also
included in figure \ref{fig:plosum}  as the solid and dashed lines
(for the even and odd modes, respectively). Over the
range of wavenumbers plotted, two normal modes are actually found
for both even and odd solutions. The two
even modes compete to be the most unstable
as $k$ varies. At $k=20$, the competition is significant,
with both modes having similar growth rates. In the initial-value
problem the competition results in pronounced 
longer-term beating oscillations in the amplitude $|A(t)|$
and $A^{-1}\partial\psi/\partial y$,
as seen in figure \ref{fig:plo2}(a,b).
In this example, the initial-value solution aligns
more closely with the slightly less unstable normal mode
(as plotted by the dashed lines),
presumably because that mode is favoured by the initial
conditions and the early transient dynamics.
A similar mode selection
is likely responsible for the discrepancy
in growth rate and phase speed between the initial-value solutions
and the generalized normal modes for $20<k<40$ and $k=5$
in figure \ref{fig:plosum}.  Otherwise, the normal-mode predictions
match the fits from the initial-value solutions and confirm
the trends of the growth rate and phase speed with wavenumber.

\subsection{Linear normal modes for base states with pseudo-plugs}
\label{sec:nm}

To conduct a more thorough normal-mode analysis, we now focus on base states
that do not develop stress discontinuities, and which, therefore, admit 
a traditional normal-mode analysis. By basing that analysis
on a matrix-based version of the linear stability problem, we
have some guarantee of finding all the unstable modes
and determining which is strongest. The situation is murkier
for the generalized normal-mode problem for discontinuous
base states discussed above, as there we proceed by continuation
from solutions without discontinuities in view of the
effective jump conditions imposed at the yield
surfaces (see \ref{sec:jumpy}).
We cannot then guarantee that the computations
locate the most unstable mode.

Figure \ref{fig:sig1plot} shows typical results for the normal-mode
problem, using two bases base states corresponding to the
stress evolutions shown in figure
\ref{fig:baspoiplot2a}(a,c). That is, for a continuous base state
with a true plug embedded within a pseudo-plug, and another
base state with an entire pseudo-plug. Both states
are unstable provided the wavenumber $k$ is taken
to be sufficiently large.
Figure \ref{fig:sig1plot} shows how the growth rate $\sigma_r$
and phase speed $c=-\sigma_i/k$ vary with wavenumber
for all the unstable modes found at $k=10$.
At this wavenumber, five unstable modes are found for
both base states.
The streamfunctions of the modes are plotted
in figure \ref{fig:sig3plot} for $k=10$ and $k=200$.
In each case, the modes have either even or odd parity,
and the gravest modes are most unstable
(matching the initial-value computations shown in
figure \ref{fig:ivpit} when $k=20$ for the base state with $a_0=0.5$).

For $k\gg1$,
the growth rates shown in figure \ref{fig:sig1plot} 
converge towards unity 
and the phase speed approaches the speed of the
pseudo-plug, $U_p$.
Simultaneously the
variations in the streamfunction $\psi(y)$
becomes localized to the pseudo-plugs
(see figure \ref{fig:sig3plot}(b)). 
These features are rationalized mathematically in
 \ref{kgg1}, where a short-wavelength analysis is
provided to extract asymptotic solutions and identify the
main balances in the equations that underscore the instability.
The analysis highlights again that key to instability are
the regions over which the base shear rate is small and
the stresses approach the yield
stress (which are now broadened into the pseudo-plugs, rather than
restricted to the yield surfaces).

Although these are not shown in figure \ref{fig:sig1plot},
further unstable modes appear at the higher wavenumbers plotted.
Most of these modes appear to detach
from a continuous spectrum that extends to
$\sigma = 0 - ik U_p$; {\it i.e.}
zero growth rate and a phase speed equal to the pseudo-plug speed $U_p$.
Further details of the spectrum are outlined in \ref{specy}.
Because the associated eigenmodes become singular,
the continuous spectrum presents challenges to the numerical calculations,
particularly in the identification of neutral modes
({\it cf.} \cite{Renardy17}).

\begin{figure}[t!]
  \centering
	\includegraphics[width=.94\linewidth]{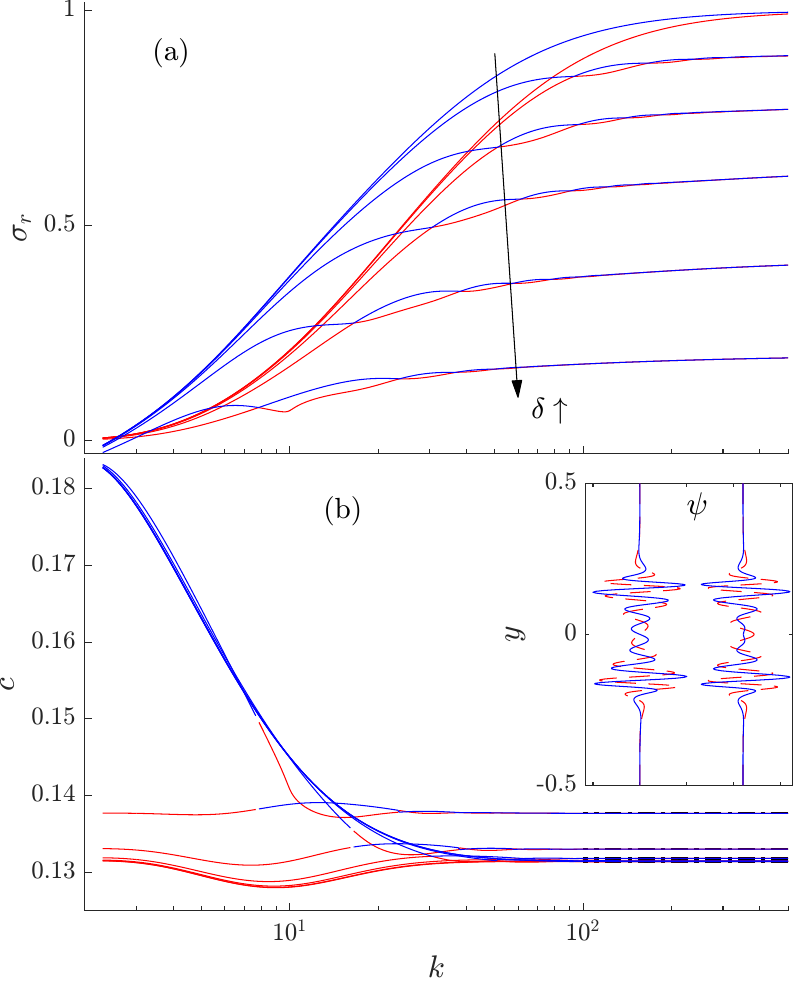}
	\caption{(a) The growth rate $\sigma_r$ and
          (b) phase speed $c=-\sigma_i/k$ of the
          two most unstable normal modes
          for $\Bi=\beta=\frac12$ with
          $\delta=\{0,0.02,0.05,0.1,0.2,0.4\}$.
          The base state has a full pseudo-plug buffered by yielded regions at the walls ($a_0=\frac12$),
          with a pseudo-plug speed $U_p$ indicated by the dot-dashed lines for $k>100$
          in (b).
          The inset show the streamfunctions of the modes with
          $(k,\delta)=(200,0.2)$ (real part in solid blue,
          imaginary part in dashed red). 
        }
	\label{fig:sig2plot}
\end{figure}

The results in figure \ref{fig:sig1plot} are computed using the
long-wave, very viscoelastic limit of the model, $\delta=0$.
Figure \ref{fig:sig2plot} extends these results to finite $\delta$,
for the base state with an entire pseudo-plug ($a_0=0.5$).
Plotted are growth rates and phase speeds against wavenumber for
the two most unstable modes, for
several values of $\delta$.
Instability becomes reduced with higher $\delta$,
but sufficiently short waves remain unstable.
The character of the short-wave modes changes, however,
as illustrated by an inset, which shows streamfunctions for
$\delta=0.2$ and $k=200$. In particular, the
solutions develop short-wavelength oscillations in
$y$ over the pseudo-plug.

\begin{figure}[t!]
    \centering
	\includegraphics[width=.9\linewidth]{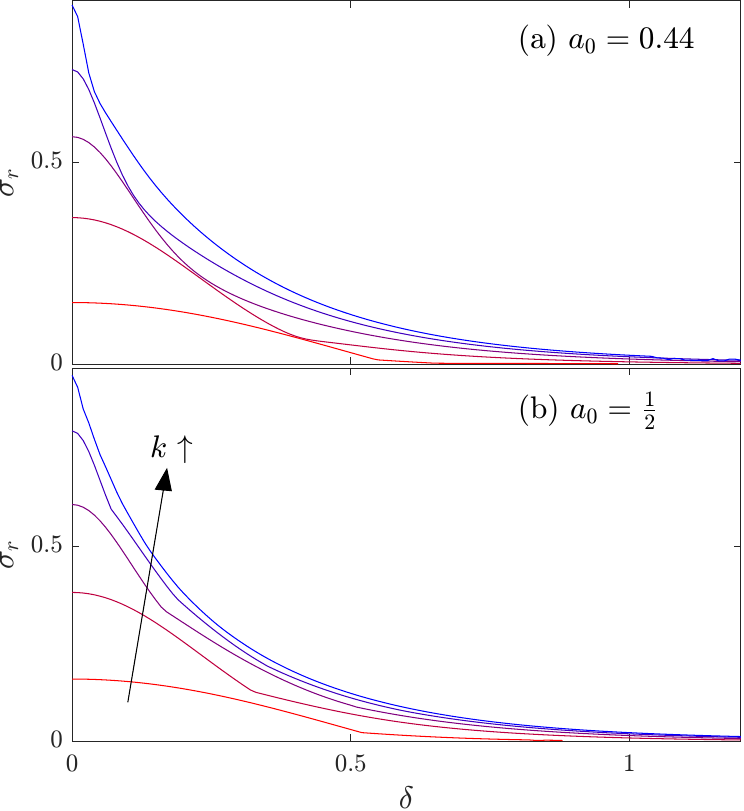}
	\caption{Growth rates $\sigma_r$
          of the most unstable mode against $\delta$ for
          $k=\{5,10,20,40,100\}$ (colour-coded, from red to blue,
          and as indicated by the arrow in (b)),
          for (a) $a_0=0.44$ and (b) $a_0=0.5$.
        }
	\label{fig:sigsdelplot}
\end{figure}

The stabilizing trend found for increasing $\delta$ is illustrated
further in figure \ref{fig:sigsdelplot}. Here, the growth rate
of the most unstable mode is plotted against $\delta$ for several
values of $k$, for the two base states with $a_0=0.44$ and $a_0=\frac12$
(the kinks in some of the curves arise from switches
in which mode is the most unstable).
Instability largely disappears once $\delta$ reaches values of
order unity, although detecting where the growth rate
vanishes is numerically challenging (in view of the
complicating presence of the continuous
spectrum; \ref{specy}). Consequently, it is difficult to establish
the critical Weissenberg number (or $\delta$) for which
the base flows become linearly stable. 

\begin{figure}
    \centering
	\includegraphics[width=\linewidth]{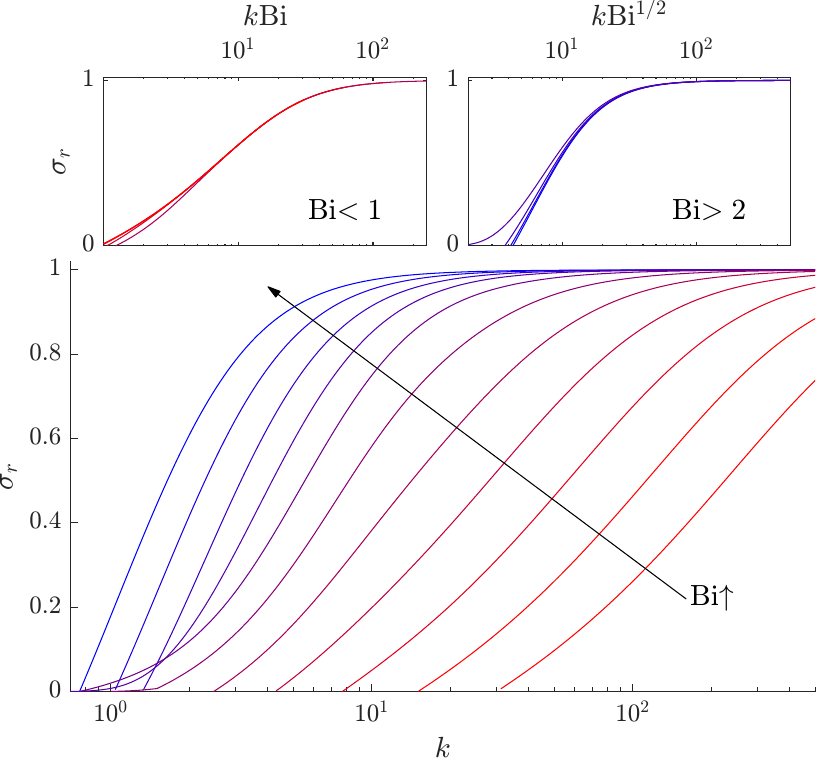}
	\caption{Growth rates $\sigma_r$
          of the most unstable mode against $k$ for
          $\Bi=\left\{\frac{1}{32},\frac{1}{16},\frac18,...,2,4,8,16,32\right\}$
          (colour-coded, from red to blue,
          and as indicated by the arrow),
          with $(a_0,\beta,\delta)=(\Bi,\frac12,0)$.
          In the upper panels, the results for $\Bi<1$ are replotted
          against $k\Bi$, and the results for $\Bi>2$ are replotted against $k\Bi^{1/2}$. 
        }
	\label{fig:sB}
\end{figure}

\subsection{Varying yield stress}

In the Oldroyd-B limit, $\Bi\to0$, base states
are linearly stable without inertia, except when
the viscosity ratio $\beta$ is very close to unity \cite{khalid21}.
Conversely, when $\Bi$ exceeds $\Bi_*$ in
\eqref{Bc} (or $(1-\beta)^{-1}$ for $\delta\ll1$),
the plugs or pseudo-plugs entirely fill the channel,
arresting the base flow. To decipher the fate of the linear instability
in both limits, we vary $\Bi$, taking $\delta=0$ for simplicity.

Figure \ref{fig:sB} plots the growth rate of the most
unstable mode against wavenumber
for a base state with a full pseudo-plug ($a_0=\Bi$)
for a range of values of $\Bi$.
When the yield stress becomes small, instability
does not disappear, but 
becomes pushed to higher $k$. Indeed, for $\Bi<1$,
the growth rates collapse close to a common curve
when plotted against $k\Bi$. This feature can be understood
by the localization of the eigenfunctions
to the pseudo-plug and the short-wavelength analysis of \ref{kgg1}.
Consequently, as $\Bi\to0$, the only instabilities that survive
are those with a spatial scale of the increasingly narrow pseudo-plug,
although their growth rates remain close to unity. 

Figure \ref{fig:sB} also demonstrates how the instability
persists even at larger yield stresses, where the
pseudo-plug fills the entire channel
(which arises when $\Bi\geq 2(1-\beta)^{-1}=4$ for $\beta=\frac12$
and $\delta=0$).
In other words, motionless base states with $\Txx=\Bi$
throughout the channel remain unstable, and, in fact, are more unstable
than moving base states. This curious result
is again predicted by the short-wavelength analysis in \ref{kgg1},
which demonstrates that instability
with $\sigma_r\to1$ still persists even when $U_p\to0$,
due to the extensional stress locked in the base state.
Moreover, the growth rates in figure \ref{fig:sB} again collapse,
but now if plotted against $k\Bi^{1/2}$ (see the top right
overlaid panel).
The result is not restricted to base states
with a full pseudo-plug, but again
applies when the centre of the channel
contains a true plug, bordered by
pseudo-plugs ({\it i.e.} to base states like that in
figure \ref{fig:baspoiplot2a}(a,b)).

\begin{figure}
    \centering
	\includegraphics[width=\linewidth]{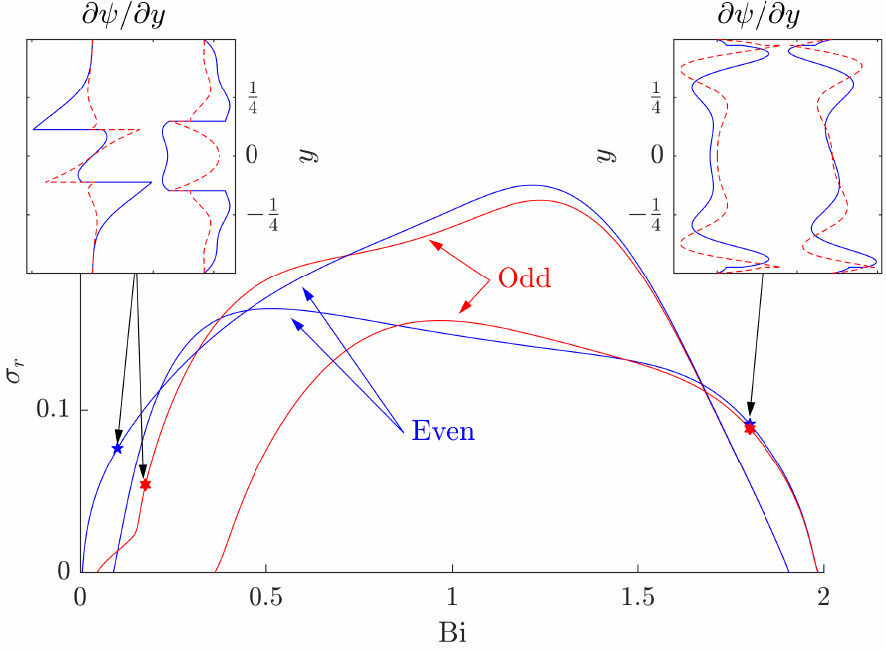}
	\caption{Growth rates $\sigma_r$ against $\Bi$
          of both unstable modes for even and odd
          perturbations for discontinous
          base states with $(k,a_0,\beta,\delta)=(20,0,\frac12,0)$.
          The overlaid panels show the eigenfunctions
          for $\partial\psi/\partial y$
          (real in solid blue, imaginary in dashed red)
          for the modes at the values of $\Bi$ indicated.
        }
	\label{fig:bitrun}
\end{figure}

To examine the situation for base states with stress discontinuities,
we continue to different yield stresses the two pairs of even and odd modes
found for $\Bi=0.5$ and $k=20$ in \S \ref{sec:ivp}.
Figure \ref{fig:bitrun} plots
the resulting growth rates as functions of $\Bi$.
As expected, the modes become stable for sufficiently small $\Bi$,
although instability looks to persist almost to the
Oldroyd-B limit for one of the modes. Unlike the base states with
pseudo-plugs, however, the instability in figure \ref{fig:bitrun}
also becomes suppressed for $\Bi \to (1-\beta)^{-1}$.
In this second limit, the yielded regions narrow to thin layers against the
walls and the decay of the
eigenfunctions across the intervening plug decouples
the perturbations on either side of channel
(see the right-hand inset to figure \ref{fig:bitrun}), as found earlier
for high $k$. The eigenvalues of the even and odd modes then
converge to one another en route to becoming damped,
as seen in figure \ref{fig:bitrun}.
Thus, in the Kevin-Voight limit of the model,
fully plugged, motionless base states
with stresses below the yield threshold are linearly stable.

\section{Discussion\label{sec:discu}}



In this paper, we have explored the start-up and stability
of pressure-driven flow down a channel for elasto-viscoplastic fluid
described by the Saramito model \cite{saramito07}. In the start-up problem,
a wide variety of steady, streamwise uniform base states can be attained,
with the initial stress conditions dictating the final structure.
When start-up commences from a stress-free initial state, flowing base states
develop normal stress discontinuities at the yield surfaces over infinitely
long times; the associated shear rate also develops a discontinuity there.
For any finite time, stresses and strain rates remain
continuous, but sharpen exponentially quickly.
On the other hand, with sufficient
extensional pre-stress, discontinuities can be avoided altogether.
Instead, base states
develop marginally yielded, plug-like flows,
or pseudo-plugs. Whilst a stress-free initial state may often be a natural
choice for computations, alternative base states may also have physical
relevance. For example, ``pre-pressuring'' by
briefly applying a strong pressure gradient to yield most of the fluid
before letting the stresses relax back to the yield stress
could potentially generate base states with pseudo-plugs.

Every base state we investigated was found to be unstable to
linear perturbations at zero Reynolds number as long as a yield stress
was present and was reached somewhere in the flow.
Instability is stronger when the base state contains a pseudo-plug.
In fact it is present even in the limit that the
yield stress is sufficient to bring the entire base flow to rest,
but the initial pre-stress yields the
entire fluid layer at the outset (so that a pseudo-plug bridges
across the whole channel).

The linear instabilities are associated with regions
in the base state where shear rates
become small. That is, either narrow regions surrounding
yield surfaces or across the pseudo-plugs. 
More awkwardly, growth rates are maximized
at the shortest streamwise wavenumbers.
These results share common features with those found
for other shear-thinning viscoelastic fluid
models or flow configurations
\cite{wilson99,bodiguel,poole16,wilson15,castillo18,patne}.
The instability is most prominent in the limit in which the
fluid is relatively visco-elastic; {\it i.e.} when Weissenberg
numbers are relatively large. As the Weissenberg
number decreases to order-one values (that is, $\delta$ increases to order-one values), the instability
becomes suppressed. This perhaps explains why
inertialess instabilities have not previously been documented
for the Saramito model. 

On a more technical note, the stress discontinuities in the base
states for the Saramito model prove awkward
in linear stability theory, as mean stress gradients appear in
the normal-mode equations. These singular coefficients reflect
how infinitesimal perturbations must correspondingly diverge in order to
shift the yield surfaces. We have avoided deriving any detailed
evolution equation for the yield surface, basing our analysis
instead on the initial-value problem for a developing base flow.
In this setting, the base-state stresses remain continuous,
although their exponential sharpening proves a notable
numerical inconvenience. We also constructed a generalized normal-mode analysis that could predict the late-time behaviour of linear perturbations in the initial-value problem. 

Nevertheless, the presence of stress discontinuities raises the question of how the yield surface
might become perturbed by a low-amplitude wave-like perturbation.
In fact, this question is significantly obscured by the
fact that to one side of the yield surface, the stress is held
at the yield stress, but on the other side the stress is held at
an order-one value above that threshold. This one-sided
aspect of the base stress implies that a low-amplitude perturbation
would only be able to shift the yield surface in one direction.
In addition to implying hysterestic motion, a further
significant mathematical complication is that 
any simple sinusoidal spatial pattern becomes ruled out.
In other words, one can no longer search for wave-like
perturbations with a single streamwise wavenumber.
Small flow disturbances must immediately become nonlinear
at the yield surfaces.

\begin{figure}[t!]
  \centering
	\includegraphics[width=\linewidth]{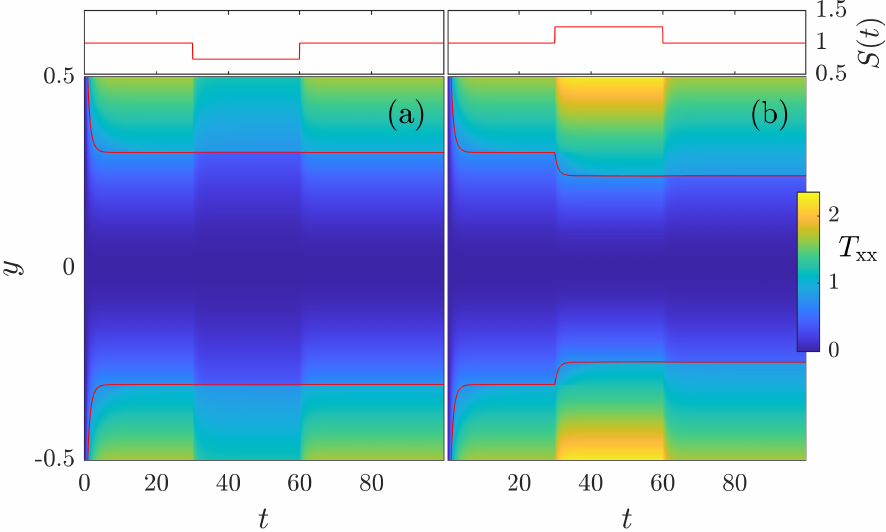}
	\caption{Base states driven by time-dependent
          pressure gradients for Saramito's model.
          For these examples, the driving terms in
          \eqref{baseq1}-\eqref{baseq2} (the $2y$ terms)
          are replaced by $2y S(t)$, where $S(t)$ is the signal
          shown in the top row of panels. The impact on the
          development of $\Txx(y,t)$ is shown in the
          density plots below; the yield surfaces are
          indicated by the red lines.)
          The parameter settings and initial conditions are
          $(\Bi,\delta,\beta)=(\frac34,0.1,\frac12)$
          and $\Txx(y,0)=\Txy(y,0)=0$.
        }
	\label{fig:absmplot}
\end{figure}

Hysteretic motion of the yield surfaces is easily illustrated
for our pressure-driven start-up flows: all we need do is introduce
a time-varying mean pressure gradient. Two examples are presented
in figure \ref{fig:absmplot}. These numerical solutions show
results in which the constant pressure gradient adopted
for the main part of our study is abruptly increased
and then decreased, or vice versa. (Practically,
this is engineered in \eqref{baseq1}-\eqref{baseq2} by replacing
$2y$ by $2y S(t)$, where $S(t)$ is a suitable signal.)

In the first example of figure \ref{fig:absmplot}(a),
the pressure gradient is suddenly decreased (at $t=30$),
then subsequently increased (at $t=60$) back to its original value.
Prior to the first switch, the flow has converged towards a base state
with stress jumps across its yield surface. One might expect that increasing the driving
pressure at this stage would lead to an abrupt thinning of the plug.
However, the yielded buffers around the plug lie at
stresses that are too high to be lowered to the yield threshold.
The yield surfaces therefore do not move. When the
driving is subsequently raised back to its original level,
the yield surfaces continue to remain at their previous positions.
By contrast in the second example of figure \ref{fig:absmplot}(b),
the driving pressure gradient is abruptly increased
(again at $t=30$) before being stepped back down. Now, the plug
is able to shrink after the first switch as the stresses
adjacent to the yield surfaces lie at the yield stress.
At the second switch however, the yield surface fails to
return to it original position. Hysteretic motion of the yield surface in shear flows described by the Saramito model has previously been noted by Burghelea \& Moyers-Gonz{\'a}lez \cite{burghelea25}.


A key focus of our analysis
has been on the long-wave, very visco-elastic
limit, where $\delta\rightarrow0$. In this limit, the Saramito model becomes dominated
by the extensional polymer stress component $\txx$,
which becomes far larger than
either the shear stress $\txy$ or the other normal
stress component $\tyy$. The physical origin of this effect stems
from the tilting of shear stresses by substantial
cross-stream shear, which builds strong extensional stresses $\txx$.
This effect is well-known in lubrication theory for Oldroyd-B-type
fluids \cite{zhang02,ahmed21,hinch24,boyko24}, and carries
over to the Saramito model. As a result,
the extensional polymer stress component $\txx$ dominates the
stress state and yield criterion, even though the deformation
rates are dominated by cross-stream shear.


To conclude, we have highlighted a number of awkward
consequences of employing Saramito's model for channel flow of an
elasto-viscoplastic fluid. Most concerning perhaps is the fact that flow
at zero Reynolds number suffers an
instability when the Weissenberg number is of order unity or larger, with the growth rate maximised at the highest streamwise wavenumbers and even
some motionless base states being unstable. 
It seems unlikely that these issues
are restricted to simple Poiseuille flow, but could well arise
in other time-dependent shear flows. At lower Weissenberg numbers, however, the instability is suppressed and so the model appears better behaved. Poiseuille channel flow could be a valuable test case for numerical methods for elasto-viscoplastic flow, challenging those methods to capture discontinuities developing around yield surfaces as well as instability at higher Weissenberg numbers. 


\section*{Acknowledgements}
JDS has been supported by a Leverhulme Trust Study Abroad Studentship. 
We thank D. M. Martinez
and M. Jalaal for helpful discussions.

\def\U{{\cal X}}
\def\V{{\cal Y}}

\appendix

\section{The sharpening steps}\label{sec:steps}

\begin{figure*}[t!]
  \centering
	\includegraphics[width=\linewidth]{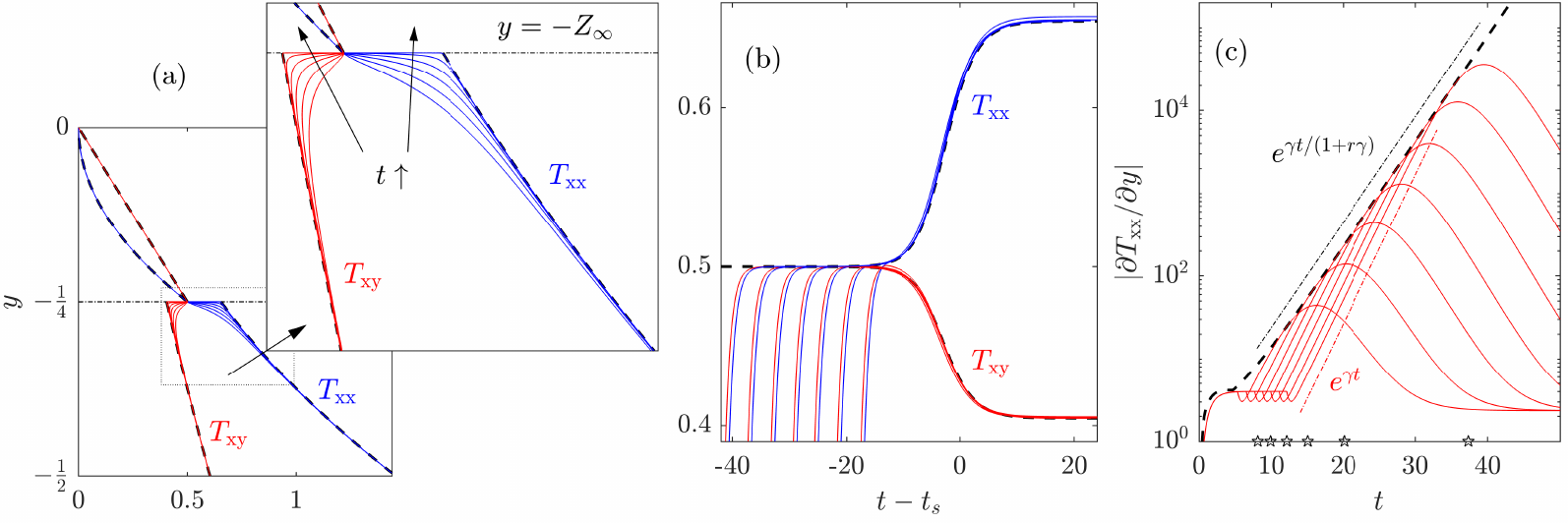}
	\caption{Base state solutions for the Saramito model,
          with $(\Bi,\delta,\beta)=(\frac12,0,\frac12)$
          and $\Txx(y,0)=\Txy(y,0)=0$.
          In (a), we plot snapshots of the stress components
          in the lower half of the channel ($y<0$).
          The yield surface is shown by the dot-dashed line,
          and the dashed lines indicate the final discontinuous
          profiles.
          Time series of $\Txx$ and $\Txy$ are plotted in (b) for
          a selection of positions near the yield surface
          ($y=-Z_\infty- 10^{-6}\times\{10^{3},400,100,40,10,4,1\}$).
          The dashed lines show the heteroclinic connection
          in \eqref{S17}.
          In (c), we plot the corresponding time series
          of $|\partial\Txx/\partial y|$.
          The dashed line shows the evolution of
          Max$(|\partial\Txx/\partial y|)$, and the dot-dashed lines show
          the exponential dependences indicated.
          The stars show the times of the snapshots plotted in (a).
        }
	\label{fig:heliplot}
\end{figure*}

In this appendix, we analyse the sharpening steps of the base states.
Figure \ref{fig:heliplot} illustrates key features of the stress
components for the base state also shown in figure \ref{fig:baspoiplot}(c,d),
but now taking $\delta=0$.
  Before any yielding takes place, the evolving stress components
  are given by \eqref{3.23}-\eqref{3.24}. If $\Txy(y,0)=0$, these
  relations can be written as
  \be
  \Txy = -2y (1-E), \qquad
  \Txx = \frac{y^2}{Z_\infty^2}\Bi (1-E)^2,
  \ee
  where
  \be
  \qquad
  E = e^{-rt},
  \qquad {\rm and} \qquad
  r = \frac{1-\beta}{\beta}
  \ee
  is another viscosity ratio.
  In the long-wave, very visco-elastic limit ($\delta\to0$),
  fluid yields when $t=t_*$ and
  \be
  E(t_*) = E_* = 1 - \frac{Z_\infty}{|y|}.
  \ee
  For $|y|$ sufficiently close to $Z_\infty$, this gives
  \be
  E_* \approx \frac{|y|-Z_\infty}{Z_\infty}
  \qquad {\rm or} \qquad
  t_* \approx \frac1r\log \left(\frac{|y|-Z_\infty}{Z_\infty}\right)^{-1}
  .
  \ee

\begin{figure*}
  \centering
	\includegraphics[width=.98\linewidth]{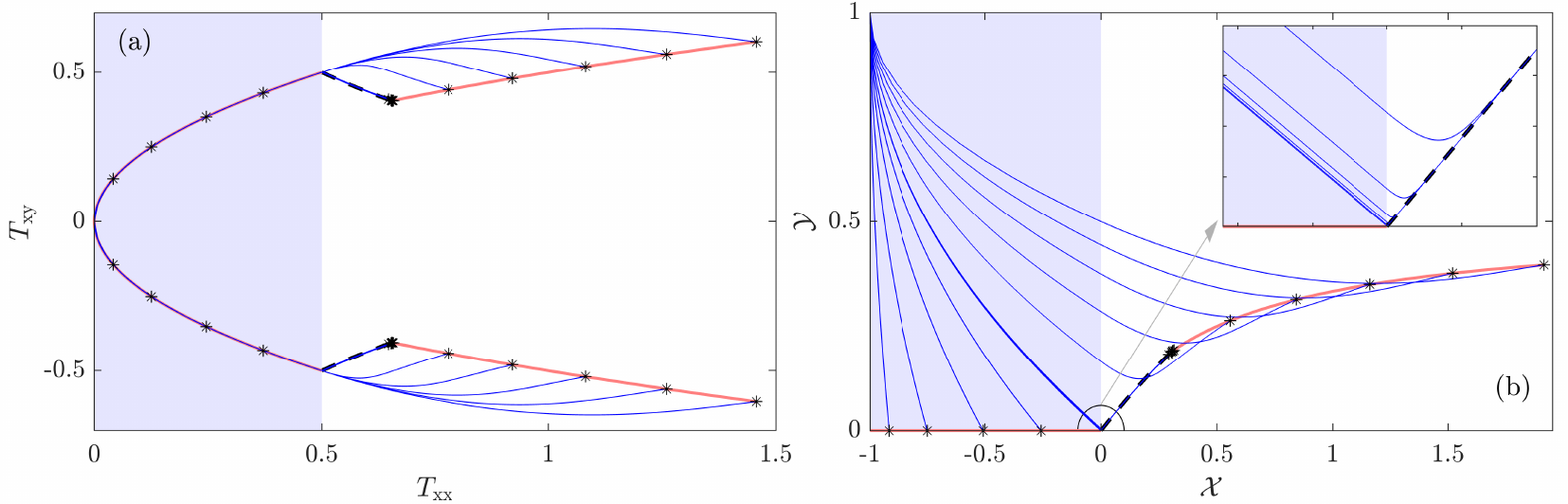}
	\caption{
          Phase portraits of the solution in
          figure \ref{fig:heliplot} on the
          (a) $(\Txx,\Txy)$ and (b) $(\U,\V)$ planes.
          The curves show the trajectories of the stress
          at the same positions as shown in
          figure \ref{fig:heliplot}, together with
          others corresponding to the stations
          $y=\pm\{0.0725,0.125,0.175,0.215,0.3,0.35,0.4,0.45,0.5\}$.
          The trajectories all begin at $(\Txx,\Txy)=(0,0)$
          or $(\U,\V)=(-1,1)$.
          The unyielded region is shaded,
          and the thicker, solid (red) lines show the loci of the final
          stress states, with the stars indicating those
          reached for the particular trajectories plotted.
          The dashed line shows the heteroclinic connection
          in \eqref{S17}.
        }
	\label{fig:hetoplot}
\end{figure*}

  Once yielding takes place
  and $t>t_*$,
  if we define the new variables
  \be
  \U(t;y) = \frac{\Txx-\Bi}{\Bi}
  \qquad {\rm and} \qquad
  \V(t;y) = \frac{2y+\Txy}{2y}
  \ee
  (where we emphasize that $y$ appears only parametrically), then the base state
  equations reduce to
  \begin{align}
    \dot{\U} & = \frac{2y^2(1-\beta)}{\beta Z_\infty^2} \V (1-\V) - \U
    \approx 2r \V (1-\V) - \U ,
    \label{S2}
    \\
    \dot{\V} &= \frac{\U(1-\V)}{1+\U} - r \V ,
    \label{S3}
\end{align}
  sufficiently close to the yield surface.
  In \eqref{S2}-\eqref{S3},
  the saddle point $\U=\V=0$ corresponds to the yield point
  of the final steady state.
  All trajectories for $|y|> Z_\infty$
  pass through to the yield point $\U=0$,
  but only those with $|y| \approx Z_\infty$ and $\V\approx0$
  pass close to the saddle and
  spend an extended period nearby;
  see figure \ref{fig:hetoplot}.

  Given that $\Txy=\Bi$ at $t=t_*$, or $\U(t_*)=0$,
  a local linearization near the saddle indicates that
  \be
  \begin{aligned}
    \U(t) &\approx a [e^{\gamma (t-t_*)} - e^{-\Gamma (t-t_*)}] ,\\
    \V(t) &\approx \frac{a}{2r}
      [(\gamma+1)e^{\gamma (t-t_*)} + (\Gamma-1)e^{-\Gamma (t-t_*)}]
      ,
  \end{aligned}
  \label{S4}
  \ee
  where
  \be
  \gamma = \frac{\sqrt{1+4\beta-4\beta^2}-1}{2\beta} ,
  \qquad 
  \Gamma = \frac{\sqrt{1+4\beta-4\beta^2}+1}{2\beta} .
  \ee
  We further have $\Txy = \mp 2 Z_\infty$ at $t=t_*$
  (for $y\approx \pm Z_\infty$),
  or
  \be
  \V(t_*) =  \frac{|y| - Z_\infty}{|y|} \approx  \frac{|y| - Z_\infty}{Z_\infty}
  .
  \ee
  Hence,
  \be
  a =
  \frac{2\beta r(|y| - Z_\infty)}{Z_\infty\sqrt{1+4\beta-4\beta^2}} =
  \frac{2\beta r E_*}{\sqrt{1+4\beta-4\beta^2}}.
  \ee

  Beyond the close passage to the saddle, the exponentially decaying
  pieces in \eqref{S4} become negligible, leaving a trajectory that
  tracks the heteroclinic connection between the saddle and 
  another fixed point of \eqref{S2}-\eqref{S3} given by
  \be
  \begin{aligned}
    \U &= \frac{\beta(\sqrt{2-2\beta+\beta^2} - 1 + \beta)}{2(1-\beta)}
    ,\\
    \V &= 1 - \half\beta - \half\sqrt{2-2\beta+\beta^2} .
  \end{aligned}
  \ee
  This heteroclinic connection is given by
  \be
  \U = \cF_\U(t-t_s)
  \qquad {\rm and} \qquad
  \V = \cF_\V(t-t_s),
  \ee
  where $\cF_\U(t-t_s)\sim e^{\gamma (t-t_s)}$ for $t\to-\infty$
  and some $t_s$ that fixes the time invariance of this solution.
  Matching this limit of the heteroclinic connection to the
  trajectory of the close passage
  to the saddle gives
  \be
  e^{\gamma (t-t_s)} = a e^{\gamma (t-t_*)} = a E_*^{r\gamma} e^{\gamma t} =
    \frac{2\beta r E_*^{1+r\gamma} e^{\gamma t}}{\sqrt{1+4\beta-4\beta^2}}.
    \label{S14}
    \ee
    Hence
    \be
    t_s = \frac{1+r\gamma}{\gamma}
    \log \left(\frac{Z_\infty}{|y|-Z_\infty}\right)
    - \frac{1}{\gamma} \log \left(
    \frac{2\beta r}{\sqrt{1+4\beta-4\beta^2}}
    \right)
    .
    \ee
    The solution post-yield is therefore given by
    \be
    \begin{aligned}
      \Txx &\approx \Bi \left[ 1 + \cF_\U(t-t_s)\right] ,\\
      \Txy &\approx \mp 2Z_\infty \left[1 - \cF_\V(t-t_s)\right].
    \end{aligned}
    \label{S17}
    \ee

    Figure \ref{fig:heliplot} compares the predictions above
    with the numerical base-state solution. Panel (a) displays
    snapshots of the stress components, magnifying the
    region in the vicinity of the yield surface to
    highlight the steepening profiles. Figure \ref{fig:heliplot}(b)
    compares time series of $\Txx$ and $\Txy$ for a selection
    of positions over this region with the heteroclinic
    connection in \eqref{S17}. For this comparison, we plot
    the stress components against $t-t_s$, which aligns
    the time series.

Note that \eqref{S17} implies that the gradient
of the extensional stress near the yield surface
is given by
\be
\pd{\Txx}{y} \approx
 \frac{(1+r\gamma)\Bi}{\gamma(|y|-Z_\infty)} \cF_\U'(t-t_s) \; {\rm sgn}(y)
 .
 \label{S19}
\ee
Very close to the saddle, we have $\cF_\U'\sim\gamma\cF$, and so
\be
\pd{\Txx}{y} \approx
\frac{(1+r\gamma)\Bi}{|y|-Z_\infty} e^{\gamma (t-t_s)} \; {\rm sgn}(y)
.
\ee
Stress gradients therefore grow exponentially with time
at fixed position, $\Txx\propto e^{\gamma t}$,
as seen in figure \ref{fig:heliplot}(c).
By contrast, the maximum stress gradients arise
further from the saddle, where $\cF_\U(T)$ reaches its maximum.
But prescribing $T=t-t_s$ in \eqref{S14} indicates that
\be
E_* \approx \frac{|y|-Z_\infty}{Z_\infty} \propto
e^{-\gamma t / (1+r\gamma)} \equiv \ve(t).
\ee
Equation \eqref{S19} therefore implies that the
maximum stress gradient has the scaling
Max$\left|\partial{\Txx}/{\partial y}\right| \propto \ve^{-1}$
({\it cf.} figure \ref{fig:heliplot}(c)).

\begin{figure}[ht!]
  \centering
	\includegraphics[width=\linewidth]{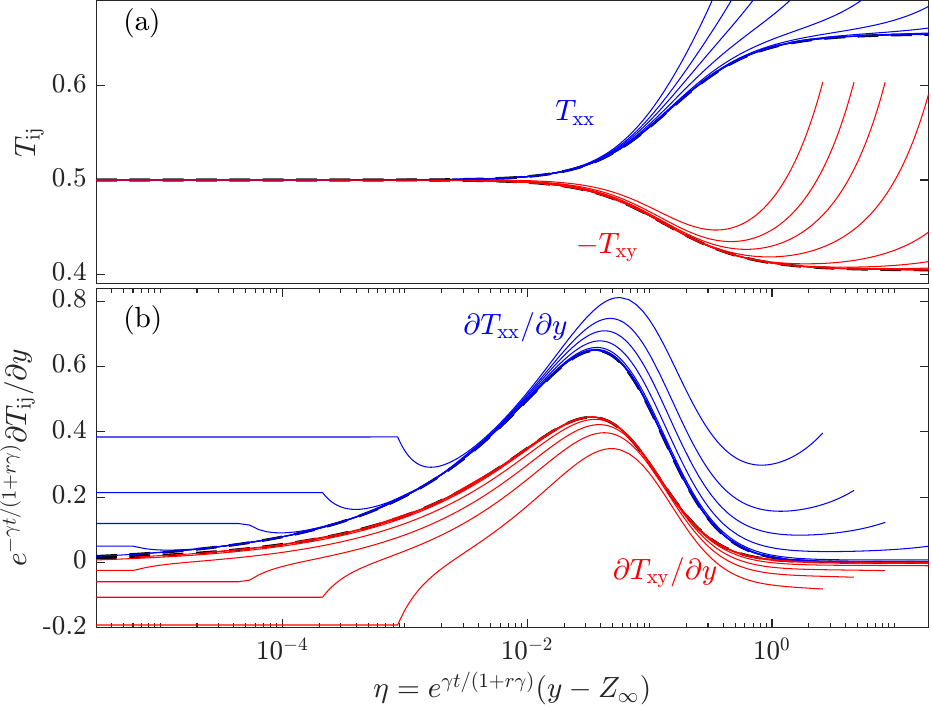}
	\caption{
          Snapshots of (a) the stress components
          and (b) their spatial gradients for
          base state solutions to the Saramito model,
          with $(\Bi,\delta,\beta)=(\frac12,0,\frac12)$
          and $\Txx(y,0)=\Txy(y,0)=0$.
          The snapshots correspond to the times shown in
          figure \ref{fig:heliplot}(c), and are plotted
          against the scaled coordinate
          $\eta = \ve^{-1}(y-Z_\infty)$.
          In (b), the stress gradients are also scaled
          by $\ve(t)$.
          The dashed lines show the
          corresponding results for the heteroclinic connection
          in \eqref{S17}.
        }
	\label{fig:hetiplot2}
\end{figure}

  Altogether, the preceding results indicate that the
  stress profiles near $y=Z_\infty$ should collapse when plotted against
  the scaled coordinate $\eta = \ve^{-1} (y-Z_\infty)$.
  Similarly, the amplification and sharpening of the
  profiles of the spatial gradients
  of the stress components can be suppressed
  by using the same scaled coordinate $\eta$ and further
  scaling $\partial T_{\rm ij}/\partial y$ by $\ve$.
  The implied collapses are illustrated in figure \ref{fig:hetiplot2},
  for the same solution of figure \ref{fig:heliplot},
  and compared with corresponding results based on
  the heteroclinic connection in \eqref{S17}.

\section{Initial-value problem for very visco-elastic long waves}\label{appA}

The stability problem simplifies slightly in the limit $\delta\ll1$.
To leading order, the model equations reduce to
\begin{align}
ik\check{p} = 
ik\ctxx &+ \ctxy' - \beta\psi''',
\qquad
\check{p}' = 0,
\label{1.16td}
\\
  \cD\ctxx
  + \Txx \Yc 
  - 2 \Ub' \ctxy
  &=  
  - ik \Txx' \psi - 
  2ik \Txx \psi' 
  - 2  \Txy \psi'' 
  \label{1.17td}
    , \\
  \cD\ctxy
  + \Txy \Yc
  - \Ub' \ctyy   
  &= - (1-\beta) \psi''
  - 
  ik\Txy' \psi 
- k^2\Txx \psi
  \label{1.18td}
  , \\
\cD\ctyy 
  = 2ik(1-\beta) \psi' 
  &- 2k^2\Txy \psi.
  \label{1.19td}
\end{align}
The perturbation equations \eqref{1.16td}-\eqref{1.19td} are
coupled to evolution equations for the base state quantities
\eqref{2.7}-\eqref{baseq2},
but with the yield condition $\cT^2=\Txx^2>\Bi^2$, which
also implies that
\begin{align}
\Yb &= \left(1-\frac{\Bi}{\Txx}\right) \Theta(\Txx-\Bi)
,
\\
\Yc &= \frac{\Bi \ctxx}{\Txx^2} \Theta(\Txx-\Bi)
\end{align}
(taking $\Txx>0$).

In practice, we solve the initial-value problem in the half domain,
$-\tfrac{1}{2}\leq y \leq0$.
Suitable integrals of \eqref{1.16td} then furnish
\begin{equation}
	\psi = \int_{-\frac12}^0(y-y')R(y',t)\,\mathrm{d}y',\label{B7}
\end{equation}
where
\begin{equation}
  \beta R = \ctxy - ik\int_{y}^0\ctxx\,\mathrm{d}y' - ik\check{p}y + A,
  \label{B8}
\end{equation}
where $A(t)$ is some function of time.
At $y=0$, we either enforce symmetry or antisymmetry boundary conditions
to search for the even or odd modes for $\psi(y,t)$,
respectively. For the even modes, we impose
\begin{equation}
	\psi'(0,t) = \check{p} = 0,
\end{equation}
\begin{equation}
	A = -2\int_{-\frac12}^0\left[\ctxy - ik\int_y^0\ctxx\,\mathrm{d}y'\right]\,\mathrm{d}y.
\end{equation}
For the odd modes, we instead set
\begin{equation}
	\psi(0,t) = \beta\psi''(0,t) + \ctxy(0,t) = A = 0,
\end{equation}
\begin{equation}
	ik\check{p} = 24\int_{-\frac12}^0y\left[\ctxy - ik\int_y^0\ctxx\,\mathrm{d}y'\right]\,\mathrm{d}y.\label{B12}
\end{equation}

To solve these equations numerically as an initial-value problem, we
first select a suitable grid in $y$. Simple quadrature formulae
are then employed to turn the integrals in
\eqref{B7}-\eqref{B12} into matrix operations.
Adopting the same spatial grid, the evolution
equations for the stress components are now integrated in time
using MATLAB's ODE45.
Practically, we use fixed but highly stretched, non-uniform spatial grids
which ensure that the mean stress gradients and the
stress perturbations all remain well resolved
over the regions adjacent to the yield surfaces ({\it cf.} \ref{sec:steps}).

\def\aa{{\cal M}}
\def\cc{{\cal N}}

\section{Normal modes for discontinuous base states}\label{sec:jumpy}

In this appendix, we derive the jump conditions that can be applied
across the sharpening steps for a base state that develops
stress discontinuities over infinite times. For simplicity,
we perform this feat in the very viscoelastic, long-wave
limit discussed in \ref{appA}.

Following the analysis of \ref{sec:steps},
we resolve the sharpening steps by introducing
the self-similar coordinate,
\be
\eta=\ve^{-1}(|y|-Z_\infty), \qquad
\ve = e^{-\gamma t/(1+r\gamma)}\ll1 ,
\ee
into the linear problem, and then searching for
local perturbations with the form,
\be
\begin{aligned}
\ctxx &\sim \ve^{-1} \vxx(\eta,t),
\\
\ctxy &\sim \ve^{-1} \vxy(\eta,t),
\\
\psi &\sim \Psi(t) + \ve \hpsi(\eta,t).
\end{aligned}
\ee
From \eqref{1.16td},
we first note that $\beta \hpsi_{\eta\eta\eta} \sim (\vxy)_\eta$,
but because the leading-order solutions
for $\vxy$ and $\hpsi_{\eta\eta}$ must decay outside the step region,
\be
\beta \hpsi_{\eta\eta} \sim \vxy .
\ee
To leading order, and
given that 
\be
\beta U' \sim -2 z_\infty -\Txy,
\qquad
z_\infty=Z_\infty\; {\rm sgn}(y)
,
\ee
the constitutive relations in \eqref{1.17td} and \eqref{1.18td}
now furnish
\begin{align} 
  \left(\hat{D} + \frac{\gamma}{1+r\gamma} + 1\right) \vxx
  + \frac{4}{\beta} (z_\infty + \Txy) \vxy
  &\sim  -ik (\Txx)_\eta  \Psi ,   \label{C18}
\\
  \left[\hat{D} + \frac{\gamma}{1+r\gamma} + \frac1\beta
    - \frac{\Bi}{\Txx}\right] \vxy
  + \frac{\Bi \Txy}{\Txx^2}\vxx
  &\sim  -ik (\Txy)_\eta  \Psi ,
  \label{C19}
\end{align}
where
\be
\hat{D} = \pd{}{t} + \frac{\gamma \eta}{1+r\gamma}\pd{}{\eta}
+ ik U_p,
\ee
and we have used $\ve_t = -\gamma \ve/(1+r\gamma)$, which helps
in the change of variables and in evaluating
$(\ve^{-1} \vxx)_t$ and $(\ve^{-1} \vxy)_t$.

\def\tc{{\check{t}}}

The forcing terms on the right
of \eqref{C18}-\eqref{C19} have the
    time dependence of $\Psi(t)$ and no other coefficients
    now depend on $t$ given that
    $\Txy=\Txy(\eta)$ and $\Txx=\Txx(\eta)$ within the
    sharpening steps.
    Hence, we may look for normal-mode-like solutions
    with dependence $e^{\sigma t}$, implying
    \be
\hat{D} = \sigma + \frac{\gamma \eta}{1+r\gamma}\pd{}{\eta}
+ ik U_p
.
\ee
Motivated by the heteroclinic solution of \ref{sec:steps},
we further switch to the new variable
\be
\tc = (\gamma^{-1}+r)\log\frac{\eta}{Z_\infty}
+ \frac{1}{\gamma}
\log \left( \frac{2\beta r}{\sqrt{1+4\beta-4\beta^2}}\right)
,
\ee
for which $\Txx=\Bi[1+\cF_\U(\tc)]$ and $\Txy=-2z_\infty[1-\cF_\V(\tc)]$,
and then put
    \be
\hat{D} = \sigma + \pd{}{\tc}
+ ik U_p
.
\ee
Equations \eqref{C18}-\eqref{C19} can then be integrated
in $\tc$ along with the
ODEs for the heteroclinic solution to find $\vxx$ and $\vxy$.
The jumps in the stress perturbations across the
steps then follow from computing
$$
\int_{z_\infty}^{z_\infty+c} (\ctxx,\ctxy) \; \di y
\sim
\int_{0}^\infty (\vxx,\vxy) \; \di\eta
$$
\be
= \frac{\gamma}{1+r\gamma}
\int_{-\infty}^\infty \eta (\vxx,\vxy) \; \di\tc
\equiv (\aa,\cc)
,
\ee
where $1\gg c\gg\ve$.

Given $(\aa,\cc)$,
the normal-mode-like solution outside the sharpening steps
can now be computed from
\eqref{1.16td}-\eqref{1.19td}, using the normal-mode
dependence $e^{\sigma t}$ and
explicitly imposing the jump conditions,
\begin{align}
  \left[\beta\psi'\right]^{z_\infty^+}_{z_\infty^-} &= \cc, \\
  \left[\beta\psi''-\ctxy\right]^{z_\infty^+}_{z_\infty^-} &= ik\aa,
\end{align}
where the $\pm$ superscripts refer to the limits on the
plugged or yielded side of $y=z_\infty$, respectively.

Practically, the jump conditions ensure that the eigenvalue problem
is not straightforward to solve by breaking the system down 
into a matrix eigenvalue equation using finite differences for spatial
derivatives. Instead, we first solve the problem in this fashion
after removing the jump conditions. The solutions can then be used
as trials for a boundary-value solver (MATLAB's bvp4c),
and the jumps gradually introduced by continuation of the solutions.
This limits the solutions that can be found directly in cases
where the modes of the problem with jumps cannot be continued
from those without jumps. This issue can be partly avoided
by continuing solutions with jumps from different wavenumbers
or other parameter settings. For $k\gg1$, it is also possible
to use an odd mode as a trial guess for an even mode, and {\it vice versa},
given the spatial localization observed for the eigensolutions.
Nevertheless, surveying the
linear stability of base states developing stress discontinuities
is more difficult
than for base states with pseudo-plugs. Similarly,
the additional terms arising in the equations for $\delta>0$
complicate matters sufficiently that we focus only on the
limit $\delta=0$ in discussing discontinuous base states.

\section{Normal modes for pseudo-plugs and $k\gg1$}\label{kgg1}

The normal-mode problem for the final steady base state solutions
without stress discontinuities
can be analyzed in the limit $k\gg1$.
We accomplish this task for the long-wave, very-visco-elastic version
of the model in \eqref{1.16td}-\eqref{1.19td}.
Thus, waves are relatively short, but not too short
(so that $1\ll k \ll \delta^{-1}$).
The normal modes have the time dependence
$e^{\sigma t}$, where
$$
\sigma =  - ikc + \varsigma , \quad
\varsigma = \sigma_0 + k^{-1} \sigma_1 + k^{-2} \sigma_2 + ... ,
$$
the leading-order phase speed is $c$ and
$\varsigma$  denotes the residual (complex) growth rate.
Because the base states has no stress discontinuities,
the central section of the channel contains
either a single pseudo-plug or a true plug buffered by
pseudo-plugs.

As long as $U-c$ does not vanish, to leading order in $k^{-1}$
\eqref{1.17td}-\eqref{1.18td} imply that
\be
\begin{aligned}
  \ctxx &\sim - 2 \Txx\left(\frac{\psi}{U-c}\right)' - \frac{\Txx'\psi}{U-c},\\
  \ctxy &\sim \frac{ik\Txx\psi}{U-c}, \label{1.32}\\
  \ctyy &\sim \frac{2ik\Txy\psi}{U-c}.
\end{aligned}
\ee
Hence, from \eqref{1.16},
\be
\ph \sim
- \Txx\left(\frac{\psi}{U-c}\right)',
\ee
or, given $\psi(-\frac12)=0$,
\be
\psi \sim \psi_I = \ph (c-U) \int_{-\half}^y \frac{\di y}{\Txx}
.
\label{bk5}
\ee
Were this solution to apply everywhere, we would be forced
into taking $c=0$ in order to satisfy the boundary condition
$\psi(\frac12)=0$. However, there is another, more interesting choice:
if $c=U_p$, where $U_p$ is the pseudo-plug speed, then
$\psi_I\to0$ on approaching the yield surface $y=y_s$, and
\eqref{bk5} can be adopted as the solution over the yielded
region $-\frac12<y<y_s$. For $y_s<y<\frac12$, we may adopt
instead
\be
\psi \sim \psi_I = -\ph (c-U) \int_y^{\half} \frac{\di y}{\Txx}
.
\ee
This exercise therefore constructs the outer pieces of an odd mode
in $\psi(y)$. Note that the
conditions $\psi'(\pm\frac12)=0$ are not satisfied by these solutions;
boundary layers of a thickness of $O(k^{1/2})$ are needed over which
this defect can be corrected.
By contrast, 
even modes in $\psi(y)$, have $\ph=0$ and vanishing a streamfunction
over the fully yielded regions.

Given $c=U_p$, \eqref{1.32} no longer apply over the 
pseudo-plugs. Here, in fact,
$\Txx=\Bi$, 
$U'=\Txx'=\Yb=0$ and $\Yc=\ctxx/\Bi$.
A different simplification of
\eqref{1.17td}-\eqref{1.18td} 
then results in
\begin{align}
  \ctxx &= \frac{2(2y\psi''-ik\Bi\psi')}{1+\varsigma},\\
  \ctxy &= -\frac{4iky\Bi\psi'}{\varsigma\Bi(1+\varsigma)}
    - \frac{k}{\varsigma}(k\Bi-2i)\psi + O(1)
  ,\\
    \ctyy &= \frac{2k[i(1-\beta)\psi' + 2ky\psi]}{\varsigma}.
\end{align}
Introducing these expressions into \eqref{1.16}
demands that
\be
  \Bi ( \sigma_0 - 1) \psi'
= 0
\label{A30}
\ee
at leading order. Hence, $\sigma_0=1$. 
Continuing further with the expansion, we find $\sigma_1=0$
and eventually arrive at
\be
2(4y^2 - \Bi) \psi'' + \sigma_2 \Bi^2 \psi
= k ( B + 2iy  \Bi \ph)
,
\label{bk10}
\ee
for some integration constant $B$. The streamfunction 
$\psi$ is therefore $O(k)$ larger than $\ph$ (and $B$)
over these regions, unlike in the fully sheared layers
where $\psi$ is the same order as $\ph$.
For the odd modes, $B=0$, whereas $\ph=0$ for the even ones.
In either case, \eqref{bk10} must be solved subject to demanding
that $(\psi,\psi')\to0$ at the edges of the pseudo-plugs
(to leading order).

The reduction to \eqref{A30} identifies the key main balances
that lead to instability in the long-wave limit when there
is a pseudo-plug: the constitutive relations
for $\ctxx$ and $\ctxy$ reduce to
\begin{align}
 \left(\pd{}{t} + ik U_p + 1\right)\ctxx  
  &\sim  2 \Txx \pd{u}{x} 
  = - 2ik \Bi \psi' 
    , \\
  \left(\pd{}{t} + ikU_p\right) \ctxy 
  &\sim  \Txx \pd{v}{x} =  - k^2\Bi \psi
  ,
  \end{align}
  which represent how viscoelastic relaxation
  of the extensional stress perturbation is driven by
  the straining of the base-state tension, and shear-stress
  perturbations become driven by cross-stream shearing of that tension.
  In combination with the leading-order streamwise force balance,
  \be
  \pd{\ctxx}{x} + \pd{\ctxy}{y} \sim 0
  ,
  \ee
  we arrive at
  \be
  k^2 \Bi  \left(\pd{}{t} + ikU_p - 1 \right) \psi' 
    \sim 0 ,
    \ee
    which is equivalent to
  \eqref{A30}. All the while,
  the relatively high wavenumber and the conditions
  met in the pseudo-plug ($U'\sim0$, $\Txx'\sim\Bi$)
  ensure that there is no stabilizing effect of any shear flow
  or base shear stress.
  
  The ODE in (D.9) can be reduced to Legendre's differential equation:
  we first eliminate the inhomogeneous terms
  by subtracting off a linear particular
solution, then define $z=2y/\sqrt{\Bi}$. After differentiating the
resulting ODE in $z$, we arrive at Legendre's ODE with
eigenvalue $\lambda=-\frac{1}{8}\sigma_2\Bi^2$, although the edges of the
pseudo-plug, $z=\pm\sqrt{1-\beta}$, are reached before the singular points at
$z=\pm1$. Regardless, we establish that
$$
\sigma_2 = - \frac{8\lambda}{\Bi^2}<0,
$$
and so
$$
\sigma \sim -ikU_p + 1 - \frac{8\lambda}{(k\Bi)^2}
.
$$
Consequently, the growth rate converges to unity
in the high wavenumber limit with a correction depending on the
combination $k\Bi$, as seen in the top left plot overlaid in figure
\ref{fig:sB}.

The preceding analysis holds even when $\Bi>(1-\beta)^{-1}$,
the pseudo-plug fills the whole channel, and $U_p=0$. The sole difference
is in the solution to (D.9), because the boundary conditions
now apply at $y=\pm1$ or $z=\pm2/\sqrt{\Bi}$, which impacts the
Sturm-Liouville eigenvalue $\lambda$. For $\Bi\gg1$, the homogeneous
solutions to (D.9) satisfy $\psi'' - \frac{1}{8}\sigma_2\Bi\psi \sim 0$,
which now implies that
$$
\sigma_2 = - \frac{8\Lambda}{\Bi},
$$
for another positive constant $\Lambda$. Hence, in this limit, we find
$$
\sigma \sim 1 - \frac{8\Lambda}{(k\sqrt\Bi)^2}
,
$$
and so a correction depending on the combination $k\sqrt\Bi$, as in
the top right plot of figure \ref{fig:sB}.

  \section{Linear spectra}\label{specy}
  
\begin{figure}
    \centering
	\includegraphics[width=.9\linewidth]{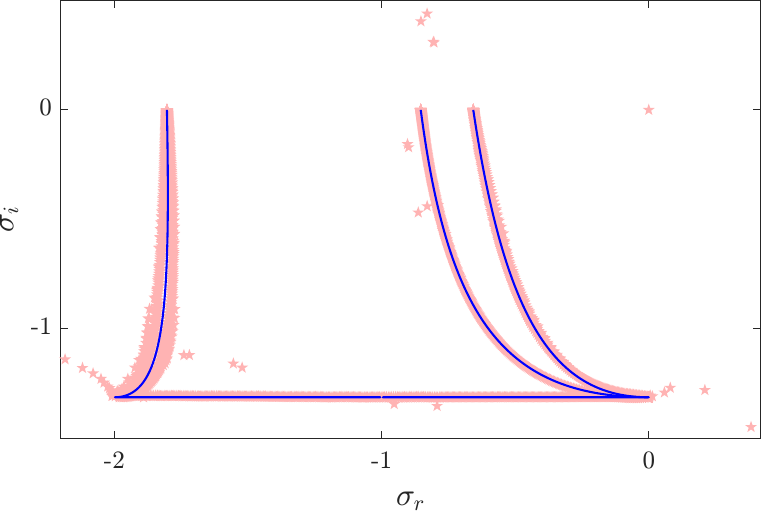}
	\caption{
          Example of the linear eigenspectrum
          for the Saramito model with
          $(k,\Bi,\delta,\beta,a_0)=(10,\frac12,0,\frac12,\frac12)$.
          The lighter (red) stars show the computed eigenvalues
          on a grid with 1000 points; the
          solid lines indicate the predictions for
          the continuous ranges of the spectrum.
}
	\label{fig:speciplot}
\end{figure}

  When the linear normal modes are computed using a finite difference
  algorithm, the results provide an approximation of the entire
  eigenspectum. As illustrated in figure \ref{fig:speciplot} for 
  Saramito's model, the spectrum contains a set of discrete
  eigenmodes together with continuous bands over the spectral
  plane ({\it cf.} \cite{Renardy17}).
  The continuous bands correspond to singular eigensolutions,
  which follow by taking $\beta\psi'' = \ctxy$, in view of \eqref{1.16},
  and then
  extracting the singular terms in $\{\ctxx,\ctxy,\ctyy\}$
  from the constitutive equations.
  We illustrate using Saramito's model with $\delta=0$:
  three bands of singular modes are found,
  given by varying position $y$ in
  \be
  \sigma = -ikU - \Yb + \left\{ \begin{array}{l}
    0 \\
    C_- \\
    C_+ 
    \end{array}\right.
  \ee
  where
  $$
  \beta C_\pm^2 +
  \left[1-\beta+\frac{\beta\Bi}{\Txx}\Theta(\Txx-\Bi)\right] C_\pm +
  $$\be
  [(1-\beta)\Txx+2\beta U'\Txy - 2 \Txy^2]\frac{\Bi}{\Txx}\Theta(\Txx-\Bi)
 = 0
  .
 \ee
A similar calculation can be made for
$\delta>0$. 
Figure \ref{fig:speciplot} shows
the approximation of the spectra
using 1000 gridpoints.
Note that, because the base states contain plugged or plug-like
regions in which $U'=\Yb=0$,
the continuous bands extend rightwards on the spectral
plane to the point $(0,-kU_p)$, where $U_p$
denotes the plug speed.
As noted in the main text, it is then possible for
unstable modes to detach from the continuous spectrum
at the bifurcation to instability.

\bibliographystyle{elsarticle-num}
\bibliography{../../vpthread/jfm.bib}

\begin{thebibliography}{10}
\expandafter\ifx\csname url\endcsname\relax
  \def\url#1{\texttt{#1}}\fi
\expandafter\ifx\csname urlprefix\endcsname\relax\def\urlprefix{URL }\fi
\expandafter\ifx\csname href\endcsname\relax
  \def\href#1#2{#2} \def\path#1{#1}\fi

\bibitem{saramito07}
P.~Saramito, A new constitutive equation for elastoviscoplastic fluid flows, J.
  Non-Newton. Fluid Mech. 145~(1) (2007) 1--14.

\bibitem{saramito09}
P.~Saramito, A new elastoviscoplastic model based on the {H}erschel--{B}ulkley
  viscoplastic model, J. Non-Newton. Fluid Mech. 158~(1-3) (2009) 154--161.

\bibitem{frag}
D.~Fraggedakis, Y.~Dimakopoulos, J.~Tsamopoulos, Yielding the yield-stress
  analysis: a study focused on the effects of elasticity on the settling of a
  single spherical particle in simple yield-stress fluids, Soft matter 12~(24)
  (2016) 5378--5401.

\bibitem{mosch}
P.~Moschopoulos, A.~Spyridakis, S.~Varchanis, Y.~Dimakopoulos, J.~Tsamopoulos,
  The concept of elasto-visco-plasticity and its application to a bubble rising
  in yield stress fluids, J. Non-Newton. Fluid Mech. 297 (2021) 104670.

\bibitem{franca2024}
H.~L. Fran{\c{c}}a, M.~Jalaal, C.~M. Oishi, Elasto-viscoplastic spreading: From
  plastocapillarity to elastocapillarity, Phys. Rev. Research 6~(1) (2024)
  013226.

\bibitem{zakeri25}
P.~Zakeri, P.~Moschopoulos, Y.~Dimakopoulos, J.~Tsamopoulos, Scaling analysis
  and self-similarity near breakup of elasto-viscoplastic liquid threads under
  creeping flow, J. Fluid Mech. 1020 (2025) A37.

\bibitem{moschopoulos23}
P.~Moschopoulos, E.~Kouni, K.~Psaraki, Y.~Dimakopoulos, J.~Tsamopoulos,
  Dynamics of elastoviscoplastic filament stretching, Soft Matter 19~(25)
  (2023) 4717--4736.

\bibitem{chaparian19}
E.~Chaparian, O.~Tammisola, An adaptive finite element method for
  elastoviscoplastic fluid flows, J. Non-Newton. Fluid Mech. 271 (2019) 104148.

\bibitem{chaparian20}
E.~Chaparian, M.~N. Ardekani, L.~Brandt, O.~Tammisola, Particle migration in
  channel flow of an elastoviscoplastic fluid, J. Non-Newton. Fluid Mech. 284
  (2020) 104376.

\bibitem{cheddadi12}
I.~Cheddadi, P.~Saramito, F.~Graner, Steady couette flows of elastoviscoplastic
  fluids are nonunique, J. Rheol. 56~(1) (2012) 213--239.

\bibitem{cheddadi08}
I.~Cheddadi, P.~Saramito, C.~Raufaste, P.~Marmottant, F.~Graner, Numerical
  modelling of foam couette flows, Eur. Phys. J. E 27~(2) (2008) 123--133.

\bibitem{walton_axial_1991}
I.~C. Walton, S.~H. Bittleston, The axial flow of a {B}ingham plastic in a
  narrow eccentric annulus, J.~Fluid Mech. 222 (1991) 39--60.

\bibitem{chen91}
K.~Chen, Interfacial instability due to elastic stratification in concentric
  coextrusion of two viscoelastic fluids, J. Non-Newton. Fluid Mech. 40~(2)
  (1991) 155--175.

\bibitem{hinch92}
E.~J. Hinch, O.~Harris, J.~M. Rallison, The instability mechanism for two
  elastic liquids being co-extruded, J. Non-Newton. Fluid Mech. 43~(2-3) (1992)
  311--324.

\bibitem{castillo22}
H.~A. Castillo, M.~R. Jovanovi{\'c}, S.~Kumar, A.~Morozov, V.~Shankar,
  G.~Subramanian, H.~J. Wilson, Understanding viscoelastic flow instabilities:
  {O}ldroyd-{B} and beyond, J. Non-Newton. Fluid Mech. 302 (2022) 104742.

\bibitem{Fielding05}
S.~Fielding, Linear instability of planar shear banded flow, Phys. Rev. Lett.
  95~(13) (2005) 134501.

\bibitem{datta22}
S.~S. Datta, A.~M. Ardekani, P.~E. Arratia, A.~N. Beris, I.~Bischofberger,
  G.~H. McKinley, J.~G. Eggers, J.~E. L{\'o}pez-Aguilar, S.~M. Fielding,
  A.~Frishman, et~al., Perspectives on viscoelastic flow instabilities and
  elastic turbulence, Phys. Rev. Fluids 7~(8) (2022) 080701.

\bibitem{wilson99}
H.~J. Wilson, J.~M. Rallison, Instability of channel flow of a shear-thinning
  {W}hite--{M}etzner fluid, J. Non-Newton. Fluid Mech. 87~(1) (1999) 75--96.

\bibitem{wilson15}
H.~J. Wilson, V.~Loridan, Linear instability of a highly shear-thinning fluid
  in channel flow, J. Non-Newton. Fluid Mech. 223 (2015) 200--208.

\bibitem{castillo18}
H.~A. Castillo, H.~J. Wilson, Elastic instabilities in pressure-driven channel
  flow of thixotropic-viscoelasto-plastic fluids, J. Non-Newton. Fluid Mech.
  261 (2018) 10--24.

\bibitem{bodiguel}
H.~Bodiguel, J.~Beaumont, A.~Machado, L.~Martinie, H.~Kellay, A.~Colin, Flow
  enhancement due to elastic turbulence in channel flows of shear thinning
  fluids, Phys. Rev. Lett. 114~(2) (2015) 028302.

\bibitem{poole16}
R.~Poole, Elastic instabilities in parallel shear flows of a viscoelastic
  shear-thinning liquid, Phys. Rev. Fluids 1~(4) (2016) 041301.

\bibitem{patne}
R.~Patne, Yield stress destabilizes viscoelastic fluid flow, Phys. Fluids
  37~(4).

\bibitem{zhang02}
Y.~L. Zhang, O.~K. Matar, R.~V. Craster, Surfactant spreading on a thin weakly
  viscoelastic film, J. Non-Newton. Fluid Mech. 105~(1) (2002) 53--78.

\bibitem{ahmed21}
H.~Ahmed, L.~Biancofiore, A new approach for modeling viscoelastic thin film
  lubrication, J. Non-Newton. Fluid Mech. 292 (2021) 104524.

\bibitem{hinch24}
J.~Hinch, E.~Boyko, H.~A. Stone, Fast flow of an {O}ldroyd-{B} model fluid
  through a narrow slowly varying contraction, J. Fluid Mech. 988 (2024) A11.

\bibitem{boyko24}
E.~Boyko, J.~Hinch, H.~A. Stone, Flow of an {O}ldroyd-{B} fluid in a slowly
  varying contraction: theoretical results for arbitrary values of deborah
  number in the ultra-dilute limit, J. Fluid Mech. 988 (2024) A10.

\bibitem{Renardy17}
Y.~Renardy, M.~Renardy, Stability of shear banded flow for a viscoelastic
  constitutive model with thixotropic yield stress behavior, J. Non-Newton.
  Fluid Mech. 244 (2017) 57--74.

\bibitem{khalid21}
M.~Khalid, V.~Shankar, G.~Subramanian, Continuous pathway between the
  elasto-inertial and elastic turbulent states in viscoelastic channel flow,
  Phys. Rev. Lett. 127~(13) (2021) 134502.

\bibitem{burghelea25}
T.~Burghelea, M.~Moyers-Gonz{\'a}lez, Elasticity mediated yielding of an
  elasto-viscoplastic fluid in a plane channel flow, Theor. Comp. Fluid Dyn.
  39~(5) (2025) 43.

\end{thebibliography}

\end{document}